\documentclass[twocolumn,preprintnumbers, amssymb,amsmath,aps,prd,nofootinbib,superscriptaddress,showpacs]{revtex4-1}

\usepackage{epsfig}
\usepackage{amssymb}
\usepackage{amsmath}
\usepackage{lipsum}
\usepackage{graphicx}
\usepackage[dvipsnames]{xcolor}
\usepackage{tikz}
\usepackage{bm}
\usepackage{subfigure}
\usepackage[
    colorlinks,
    linkcolor=blue,
    anchorcolor=black,
    citecolor=blue
]{hyperref}
\usepackage{soul}
\usepackage[version=4]{mhchem}
\usepackage{multirow}
\usepackage{makecell}
\usepackage{array}
\allowdisplaybreaks[4]

\newcommand\pperp{\ensuremath{P_J}}

\newcommand\dd{\ensuremath{\mathrm{d}}}

\begin{document}

\title{Extracting Essential Non-perturbative Information in Jet Invariant Mass via the Bayesian Analysis}

\author{Zhan Gao} \email{gaozhan2@mails.ccnu.edu.cn}
\affiliation{Key Laboratory of Quark and Lepton Physics (MOE) and Institute of Particle Physics, Central China Normal University, Wuhan 430079, China}

\author{Yu Shi} 
\email{yu.shi@polytechnique.edu} 
\affiliation{Key Laboratory of Particle Physics and Particle Irradiation (MOE), Institute of frontier and interdisciplinary science, Shandong University, Qingdao, Shandong 266237, China}
\affiliation{CPHT, CNRS, \'Ecole polytechnique,  Institut Polytechnique de Paris, 91120 Palaiseau, France}

\author{Bo-Wen Xiao}  
\email{xiaobowen@cuhk.edu.cn}
\affiliation{School of Science and Engineering, The Chinese University of Hong Kong, Shenzhen, Guangdong, 518172, P.R. China}
\affiliation{Southern Center for Nuclear-Science Theory (SCNT), Institute of Modern Physics, Chinese Academy of Sciences, Huizhou 516000, Guangdong Province, China}

\author{Han-Zhong Zhang} \email{ zhanghz@mail.ccnu.edu.cn }
\affiliation{Key Laboratory of Quark and Lepton Physics (MOE) and Institute of Particle Physics, Central China Normal University, Wuhan 430079, China}

\begin{abstract}
In this paper, we present a new three-dimensional non-perturbative (NP) function to account for and parameterize the NP contributions in the jet invariant mass spectrum, in addition to the conventional NP mass shift parametrization. By implementing Bayesian analysis on experimental data of the jet invariant mass from exclusive $W/Z+$jet events and inclusive jet events in $pp$ collisions at RHIC and LHC, where collisional energy increases by a factor of up to $65$ from RHIC to LHC, we ensure the analysis covers a wide range of data. For the first time, we simultaneously extract NP contributions from hadronization, initial soft-gluon radiation, and underlying events, based on two different NP prescriptions. We find that the contribution from initial soft-gluon radiation is negligible, and the hadronization effect dominates in the small-$R$ region, while underlying events provide the dominant contribution in the large-$R$ region. Moreover, when only hadronization effects are considered, our results successfully describe the jet mass data measured in $e^+e^-$ collisions, where only hadronization effects are expected to be present. Our work offers quantitative insights into understanding the soft hadronic contribution to jet substructure.
\end{abstract}
\maketitle

\textit{Introduction---} 
Jets and jet substructure have served as powerful probes for high-precision tests of Quantum Chromodynamics (QCD)~\cite{Pumplin:2009nk,Watt:2013oha,ATLAS:2013pbc,CMS:2014qtp,CMS:2016lna,Currie:2016bfm,Harland-Lang:2017ytb,Currie:2017eqf,Marzani:2019evv,Gutierrez-Reyes:2019msa,AbdulKhalek:2020jut,ATLAS:2021qnl,CMS:2021iwu,ALICE:2021njq,Benitez:2024nav,CMS:2013vbb, CMS:2014mna, ATLAS:2015yaa, ATLAS:2017qir, Britzger:2017maj,Hannesdottir:2022rsl,Benitez:2025vsp} 
 and for searching for new physics beyond the Standard Model~\cite{Soper:2010xk,Godbole:2014cfa,Chen:2014dma,Adams:2015hiv,ATLAS:2019fgd,CMS:2020cpy,CMS:2024nsz,CMS:2025wfw} in the past decade. Moreover, jet measurements have become important tools for constraining the strong coupling constant~\cite{CMS:2013vbb, CMS:2014mna, ATLAS:2015yaa, ATLAS:2017qir, Britzger:2017maj,Hannesdottir:2022rsl,Benitez:2025vsp} in the small collisional systems, and for probing properties of the quark-gluon plasma (QGP) in heavy-ion $AA$ collisions~\cite{Mangano:2017plv,Milhano:2017nzm,Chang:2017gkt,KunnawalkamElayavalli:2017hxo,Ringer:2019rfk,Casalderrey-Solana:2019ubu,Caucal:2019uvr,Caucal:2021cfb,Cunqueiro:2023vxl,JETSCAPE:2023hqn,Chien:2024uax,Apolinario:2024equ}. Achieving these goals requires high-precision perturbative calculations incorporating higher-order corrections, the resummation of large logarithms, and non-perturbative (NP) models accounting for various dominant NP contributions. Jet-related observables are particularly sensitive to NP soft physics, such as hadronization, soft gluon radiation (ISR), as well as underlying event (UE) and pile-up effects~\cite {Moraes:2007rq,Cacciari:2007fd}. Consequently, understanding leading NP soft contributions remains essential in high-energy collisions. However, due to the complexity of the NP background, these effects are challenging to address clearly within a perturbative framework. 

The jet invariant mass, one of the most fundamental and simplest observables in the jet substructure, serves as a key probe for distinguishing electroweak signals from QCD backgrounds and has attracted significant interest from both theoretical and experimental research~\cite{Larkoski:2017jix,Kogler:2018hem,Marzani:2019hun}. In recent years,  a large amount of data has been collected across a wide range of colliding systems from LEP~\cite{Chien:2010kc, Chen:2021uws}, HERA~\cite{H1:2024pvu} to Tevatron~\cite{CDF:2011loy}, RHIC~\cite{STAR:2021lvw} and the LHC~\cite{ATLAS:2012am,CMS:2013kfv, ALICE:2017nij,CMS:2017pcy,ATLAS:2017zda, ATLAS:2018jsv, CMS:2018fof, CMS:2018vzn, ALICE:2024jtb, LHCb:2025tvf}. Perturbative QCD has enabled substantial theoretical progress in the study of jet mass distributions. In the large-mass region, higher-order corrections play a dominant role and provide a sufficient framework for describing the jet mass measurements~\cite{Gehrmann-DeRidder:2007vsv,Weinzierl:2009ms,Dasgupta:2012hg,Frye:2016okc,DelDuca:2016ily,Idilbi:2016hoa,Baron:2018nfz,Kardos:2018kth,Kardos:2020gty,Ziani:2021dxr}. In the small-mass region, various types of large logarithms hinder the convergence of fixed-order perturbative expansions, requiring resummation techniques to restore the perturbative predictive power~\cite{Catani:1990rr,Catani:1993hr,Banfi:2004yd,Dasgupta:2012hg,Li:2012bw,Chien:2012ur,Frye:2016aiz,Idilbi:2016hoa,Frye:2016okc,Kolodrubetz:2016dzb,Marzani:2017kqd,Kang:2018jwa,Balsiger:2019tne,Kardos:2020gty,Ziani:2021dxr,Gaid:2024tie,Hoang:2025uaa, Dasgupta:2001sh,Dasgupta:2002bw,DuranDelgado:2011tp,Khelifa-Kerfa:2011quw,Schwartz:2014wha,Larkoski:2015zka,Larkoski:2016zzc,Becher:2016mmh,Becher:2017nof,Caletti:2021oor,Becher:2021zkk,Becher:2023vrh,Becher:2023mtx}. Furthermore, NP effects in the small mass region become considerably important. For instance, hadronization corrections have been taken into account in several ways: by modeling parton-to-hadron transverse momentum transitions~\cite{Korchemsky:1999kt,Dasgupta:2007wa,Li:2012bw, Jouttenus:2013hs,Stewart:2014nna}, by modeling parton-to-hadron angular modifications~\cite{Ringer:2019rfk}, and constructing parton-to-hadron level transfer matrices in event generators~\cite{Reichelt:2021svh}.

To mitigate NP contributions, a series of jet grooming techniques have been developed to remove wide-angle radiations within jets~\cite{Ellis:2009su,Ellis:2009me,Krohn:2009th,Dasgupta:2013ihk,Larkoski:2014wba}. While these widely used jet grooming techniques can partially suppress NP effects, significant NP effects remain~\cite{Dasgupta:2013ihk,Dasgupta:2015lxh,Marzani:2017mva,Marzani:2017kqd,Hoang:2017kmk,Kang:2018jwa,Kang:2018vgn,Hoang:2019ceu,Ferdinand:2023vaf,Pathak:2023sgi,Dhani:2024gtx}. Thus, conducting a global analysis of these NP effects is crucial for performing high-precision QCD tests with jet-related observables.

Traditionally, NP effects on the jet invariant mass have been accounted for using the momentum shift prescription~\cite{Stewart:2014nna} (referred to as the mass shift prescription for short), which introduces NP shift contributions to the jet mass. Inspired by Ref.~\cite{Ringer:2019rfk}, we propose a novel NP model that incorporates angular splitting in three-dimensional space during the hadronization process, providing a complementary description of the jet invariant mass observable. To explore the implications of both prescriptions, we perform a Bayesian analysis of high-precision experimental data at next-to-leading logarithmic (NLL) accuracy. The dataset includes $W/Z+$jet events~\cite{CMS:2013kfv} and inclusive jet events in $pp$ collisions, with data covering experiments from RHIC~\cite{STAR:2021lvw} to the LHC~\cite{ATLAS:2012am, ATLAS:2018jsv}. This analysis provides fits to the data using both the momentum shift prescription and the newly proposed three-dimensional angular prescription (referred to as the $\theta$ prescription for short), enabling the extraction of NP contributions from hadronization, ISR, and UE to the jet invariant mass. As an independent validation, we apply both NP prescriptions to describe $e^+e^-$ data~\cite{Chen:2021uws}, which involve only hadronization effects. This verification demonstrates the robustness of our framework across different experimental environments.

This paper is organized as follows: First, we introduce the NP parametrizations for jet mass and then provide the essential details about the Bayesian analysis used in fitting all the relevant datasets. Before concluding, we present a sample of the numerical results while leaving all detailed comparisons in the supplemental material.

\textit{Two NP Parametrizations for Jet Mass---} Let us first present a parametrization that incorporates the three-dimensional NP effect. In the small-$R$ limit for a hard jet with transverse momentum $P_J$, the invariant jet mass arising from a parton radiating a gluon is given by $m_0^2 = \xi(1-\xi)P_J^2\theta_0^2$, where $\xi$ denotes the longitudinal momentum fraction of the parton, and $\theta_0$ represents the opening angle between the parton and the radiated gluon.

\begin{figure}[!h]
\includegraphics[width=0.8\linewidth]{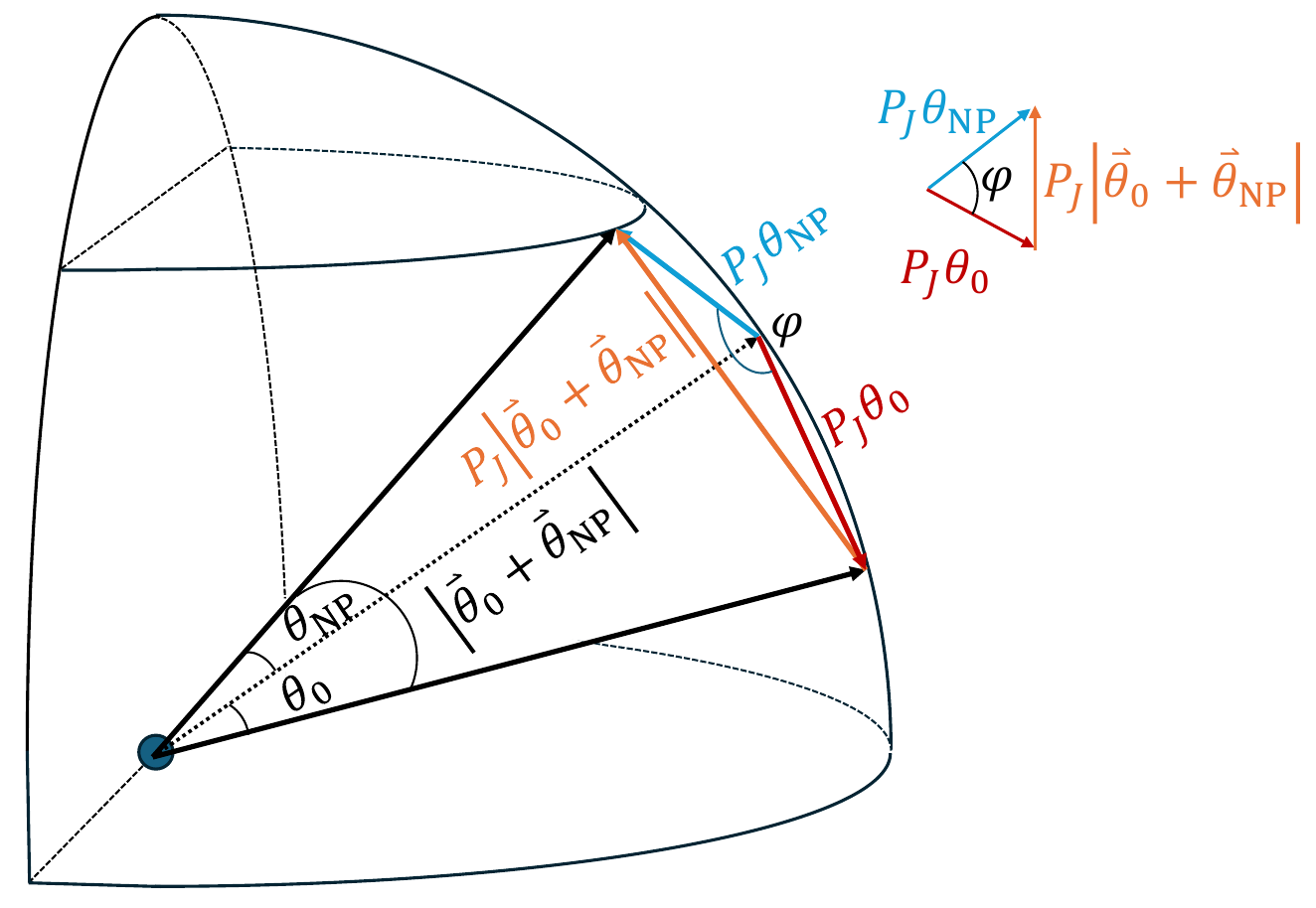}
\caption{Illustration of the combined jet mass contributions from NP hadronization effects and the perturbative part.}
\label{fig:NP_cartoon1}
\end{figure}

Following the NP prescription for jet substructure used in Ref.~\cite{Ringer:2019rfk}, the NP hadronization mechanism can ``kick'' the final hadron away from the direction of the original parton by an additional angular deflection $\vec{\theta}_{\rm NP}$. As illustrated in Fig.~\ref{fig:NP_cartoon1}, this deflection occurs in three-dimensional space. Denoting $P_J\vec{\theta}_0$ and $P_J\vec{\theta}_{\rm NP}$ as the three-dimensional vectors in the small-angle (base-length) approximation, the jet invariant mass, including the NP hadronization correction, becomes:
\begin{eqnarray}
m^{2} &=&  \xi(1-\xi)  P_{J}^{2}\left(\vec \theta_0+\vec  \theta_{\rm NP}\right)^{2} \nonumber \\
&=& \xi(1-\xi)  P_{J}^{2} \left( \theta_0 ^{2}+2   \theta_0 \theta_{\rm NP} \cos\phi+  \theta_{\rm NP}^2 \right) , 
\end{eqnarray}
where $\phi$ is the angular difference between $P_J\vec{\theta}_0$ and $P_J\vec{\theta}_{\rm NP}$. Equivalently, if we define the NP jet mass as $m_{\rm NP}^2 = \xi(1-\xi)P_J^2\theta_{\rm NP}^2$, then it is straightforward to find that the invariant mass with NP corrections can be written as:
\begin{equation}
m^{2} = m_0^{2} + 2 m_0 m_{\rm NP} \cos\phi + m_{\rm NP}^2. 
\end{equation}
where $\phi$ represents the directional shift caused by hadronization in three-dimensional space. This expression implies that NP effects can be three-dimensional and generate an angular mass shift as shown in Fig.~\ref{fig:NP_cartoon1}.

When a parton converts into a hadron and another particle, which usually carry nonzero transverse momentum during the splitting, a non-trivial jet mass can arise from this NP effect. To model the NP effect due to hadronization, we assume that it generates a typical transverse momentum broadening of the order $p_\perp^2$, thus it gives
\begin{equation}
m_{\rm NP}^2 = \frac{p_\perp^2}{z(1-z)}= \frac{p_\perp^2 P_J^2 R^2}{z(1-z) P_J^2 R^2} 
\propto
\frac{p_\perp^2  R^2 P_J^2}{m^2}.
\end{equation} 
For simplicity, we assume that the distribution of this NP jet mass contribution is described by a simple Gaussian NP function $G(m_{\rm NP})$, which satisfies the following Gaussian distribution
\begin{equation}
G(m_{\rm NP})=\frac{1}{\pi\lambda^{2}}e^{-\frac{m_{\rm NP}^{2}}{\lambda^{2}}},
\end{equation}
with the NP mass width $\lambda$ can be parametrized as $\lambda^2 =Q^2_hP^2_J/m^2$ with
$Q_h^2=a^2_h R^2$. By implementing the NP angle prescription, the invariant mass distribution involving the NP contribution can be written as
\begin{eqnarray}
&& \frac{d\sigma}{dm} =\int d m_0 d\phi  d m_{\rm NP} m_{\rm NP} \frac{d\sigma}{dm_{0}}G(m_{\rm NP})\delta\left(m-f(m_0)\right)
 \nonumber \\
&&=
\int   d\phi d  m_{\rm NP} m_{\rm NP} \frac{d\sigma}{dm_{0}}\frac{  m G(m_{\rm NP})}{\sqrt{m^{2}-m_{\rm NP}^{2}(1-\cos ^2\phi)} }  ,
\label{eq:np1}
\end{eqnarray}
with $f(m_0)= \sqrt{ m_0 ^{2}+2 m_0 m_{\rm NP}  \cos\phi+m_{\rm NP}^{2}}$ and $m_{0}=- m_{\rm NP} \cos \phi \pm
\sqrt{m^{2}-m_{\rm NP}^{2}(1-\cos ^2\phi)} \geq 0 $. It is important to note that, in our framework, the $R$ dependence of the hadronization contribution is proportional to even powers of the jet cone size $R$, which differs from the NP behavior proportional to odd powers of the jet cone size $R$ found in Ref.~\cite{Stewart:2014nna}. The latter encapsulates the simple shift in the jet mass spectrum due to hadronization effects, characterized by a dependence on odd powers of the jet radius $R$ and universality for quark and gluon jets at small $R$. Consequently, these dimensional differences in the treatment of variables lead to distinct behaviors of NP effects within each approach.

Based on the above parametrization, we further extend the NP parameter $Q_h^2$ to include other significant contributions from ISR and UE as follows: $Q_{\rm NP}^{2} = Q_{h}^{2} + Q_{i}^{2} + Q_{\rm UE}^{2}$, 
where $Q_h^2$ represents the hadronization effects in the final state, $Q_i^2$ accounts for ISR before the hard collision, and $Q_{\rm UE}^2$ captures soft UE effects that increase with the collision energy $\sqrt{s}$. For the dependence on cone size and energy, we adopt the following ansatz: $Q^2_{i} =(a^2_{i,a} + a^2_{i,b})R^4$ GeV$^2$, $Q^2_{h} = a^2_{h,c} R^2$ GeV$^2$, and $Q^2_{\rm UE} = a^2_{\rm UE} \sqrt{s} R^4$ GeV$^2$, where the dimensionless NP coefficients $a_{i,a}$, $a_{i,b}$, $a_{h,c}$, and $a_{\rm UE}$ are constant parameters to be determined from experimental data with $a,b,c=q,g$. $\sqrt{s}$ is a number given in the unit of TeV.

In addition to the three-dimensional NP prescription proposed above, there has been a popular NP function used to capture all hadronization, ISR, and UE effects based on a momentum shift (mass shift) prescription\cite{Stewart:2014nna}. In this case, the modified jet mass distribution is given by
\begin{equation}
\frac{d\sigma}{dm} = \int d k_t \frac{d\sigma}{dm_0} F_k(k_t)
\end{equation}
with $m^2_0=m^2-2 k_t P_J$. The shape function likes $F_k(k_t) =\frac{4k_t }{\Omega_{\rm NP}} e^{-\frac{2k_t}{\Omega_{\rm NP}}}$. The NP factor includes hadronization, ISR, and UE contributions, which are given as
$\Omega_{\rm NP} = \Omega_{ h} + \Omega_{ i} + \Omega_{\rm UE}$ with the corresponding parametrization form is given as $\Omega_{ i} =(b_{ i,a} +b_{i,b}  )R^4$ GeV, $\Omega_{ h}=b^{(1)}_{ h,c} R+b^{(3)}_{ h,c} R^3$ GeV,  and $\Omega_{\rm UE} =b_{\rm UE} \sqrt{s} R^4$ GeV.

 To relate quark and gluon NP parameters and account for their color factor difference, we impose $a_{h/i,q}^2=\frac{C_F}{C_A} a_{h/i,g}^2$ in the $\theta$ prescription and $ b_{h/i,q}= \frac{C_F}{C_A}  b_{h/i,g}$ in the mass shift prescription. We further assume that the underlying-event contribution is flavor‐independent, i.e., identical for quarks and gluons. Details of the theoretical framework are provided in the Supplemental Material.

\textit{Bayesian analysis---} We use Bayesian analysis to extract quantitative NP information for jet mass by combining theoretical predictions with all relevant experimental data. Bayesian analysis is a statistical tool that combines prior assumptions with observed data to update posterior beliefs about unknown parameters in a probabilistic manner. This method allows us to infer the most likely NP parameters and their uncertainties from experimental data. The core idea is expressed by Bayes' theorem $\mathcal{P}(\boldsymbol{x} |\textbf{data}) \propto \mathcal{P}(\boldsymbol{x}) \mathcal{P}(\textbf{data}|\boldsymbol{x})$, which relates the prior distribution $\mathcal{P}(\boldsymbol{x})$ (representing initial assumptions) and the likelihood function $\mathcal{P}(\textbf{data}|\boldsymbol{x})$ (capturing the information provided by the data) to form the posterior distribution $\mathcal{P}(\boldsymbol{x} |\textbf{data})$, which contains updated knowledge after incorporating the data. Here, $\boldsymbol{x}$ represents the parameter set, and $\textbf{data}$ denotes the experimental data.

In our analysis, we employ the NLO CTEQ PDFs~\cite{Hou:2019efy} and utilize the resummed improved cross-section at NLL accuracy. To improve computational efficiency and reduce resource usage, we employ the Gaussian Process (GP) emulator~\cite{GPE1,GPE2} for rapid and accurate model evaluations.  The GP emulator is trained on a set of $500$ randomly sampled points through the theoretical model. We take that the prior distribution $\mathcal{P}(\boldsymbol{x})$ to be uniform over the region $\boldsymbol x \in [0,5]$. Next, We employ the Markov Chain Monte Carlo (MCMC) algorithms~\cite{Goodman:2010dyf,Foreman-Mackey:2012any}, utilizing one of the most widely used Metropolis-Hastings algorithms to iteratively sample the posterior distribution. The full chain consists of $10^4$ steps, with the final $5000$ steps retained for analysis to ensure equilibration and minimize biases from initial transients.  

The experimental data analyzed in this study include measurements of single-inclusive jet and $W/Z$-tagged exclusive jet mass in $pp$ collisions. The single-inclusive jet mass data are sourced from various experiments: STAR data at $\sqrt{s} = 200~\text{GeV}$~\cite{STAR:2021lvw},  and ATLAS data at $\sqrt{s} = 7~\text{TeV}$~\cite{ATLAS:2012am} and $\sqrt{s} = 5.02~\text{TeV}$~\cite{ATLAS:2018jsv}. Measurements of $W/Z$-tagged exclusive jet mass are based on CMS data at $\sqrt{s} = 7~\text{TeV}$~\cite{CMS:2013kfv}. 
Furthermore, as the robustness check, we investigate the impact of the CMS measurements of dijet jet mass at  $\sqrt{s} = 13~\text{TeV}$~\cite{CMS:2018vzn} on the NP values extracted via Bayesian inference using these two NP prescriptions. The full analysis is presented in the Supplemental Material.

In order to focus on the low jet mass regime, we restrict both theoretical predictions and experimental inputs to $m\ll P_JR$. Furthermore, to assess the robustness of our extraction, we conduct the analysis using two independent small-mass data sets. Details can be found in the Supplemental Material.

\begin{figure}[ht]
\includegraphics[width=0.99\linewidth]{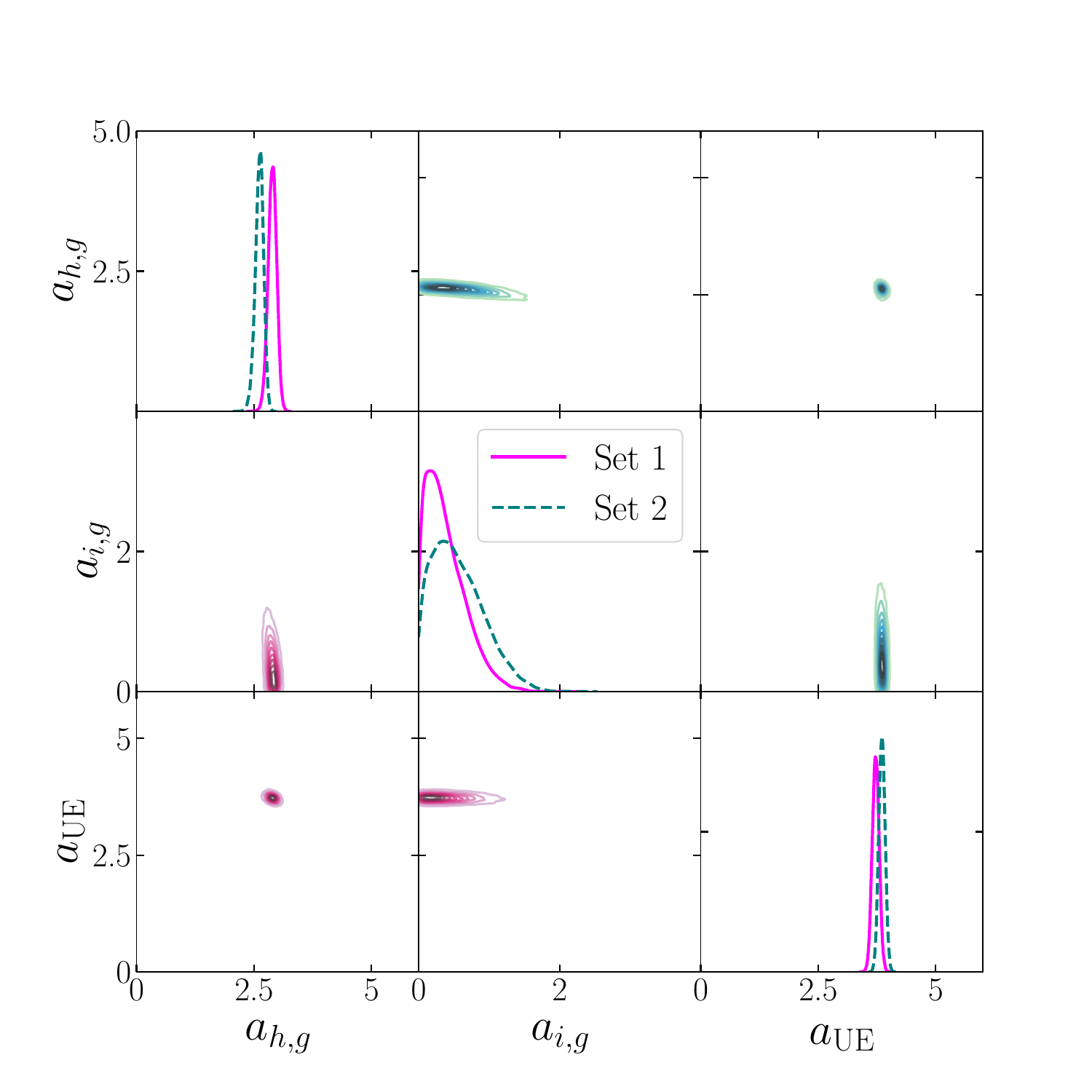}
\caption{Posterior distributions and correlations of the gluon parameters in the $\theta$ prescription.}
\label{fig:theta_Bayes}
\end{figure}

\begin{table}[!ht]
\begin{tabular}{|c|c|c|c| }
\hline
\multirow{2}{*}{Parameter}  & \multirow{2}{*}{Prior range} & \multicolumn{2}{c|}{MAP}  \\\cline{3-4} 
          &               & Set 1         & Set 2         \\\hline
$a_{h,g}$     & $[0, 5]$ & $2.907_{-0.169}^{+0.127}$ & $2.637_{-0.178}^{+0.113}$  \\\hline
$a_{i,g}$     & $[0, 5]$ & $0.180_{-0.146}^{+0.745}$ & $0.324_{-0.259}^{+0.911}$  \\\hline
$a_{\rm UE}$  & $[0, 5]$ & $3.712_{-0.107}^{+0.129}$ & $3.856_{-0.109}^{+0.108}$  \\\hline
\end{tabular}
\caption{The table provides the NP parameters, along with their prior ranges and the constrained maximum likelihood values, including uncertainty estimates at the $90\%$ credible intervals within the $\theta$ prescription.  }
\label{tab:theta}
\end{table}

\begin{figure}[ht]
\includegraphics[width=0.99\linewidth]{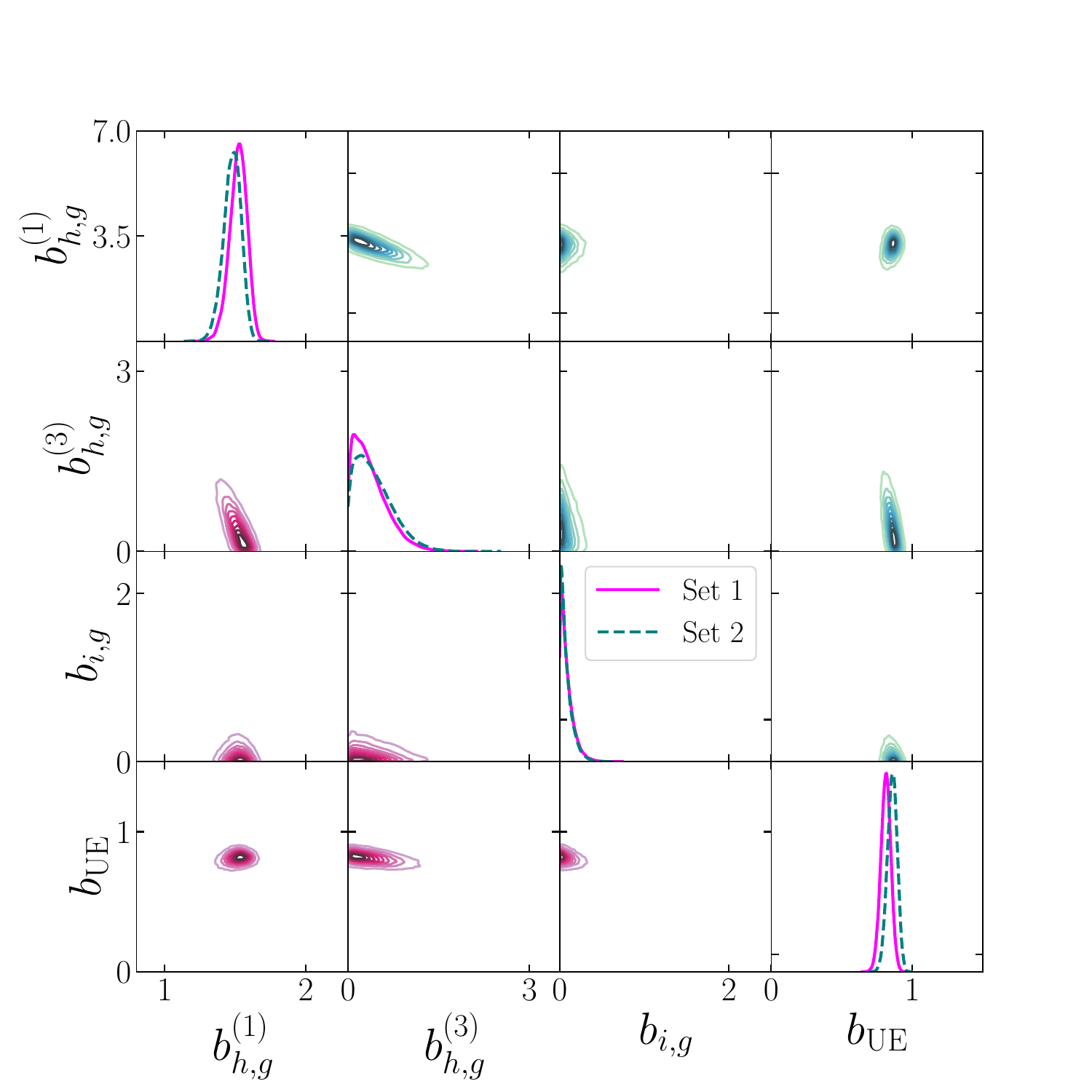}
\caption{Posterior distributions and correlations of parameters in the mass shift prescription.}
\label{fig:kt_Bayes}
\end{figure}

\begin{table}[!ht]
\begin{tabular}{|c|c|c|c| }
\hline
\multirow{2}{*}{Parameter}  & \multirow{2}{*}{Prior range} & \multicolumn{2}{c|}{MAP}  \\\cline{3-4} 
          &             & Set 1   & Set 2 \\\hline
$b_{h,g}^{(1)}$ & $[0, 5]$ & $1.530_{-0.116}^{+0.086}$ & $1.494_{-0.125}^{+0.080}$ \\\hline
$b_{h,g}^{(3)}$ & $[0, 5]$ & $0.084_{-0.055}^{+0.829}$ & $0.224_{-0.185}^{+0.800}$ \\\hline
$b_{i,g}$       & $[0, 5]$ & $0.018_{-0.013}^{+0.222}$ & $0.018_{-0.013}^{+0.206}$ \\\hline
$b_{\rm UE}$    & $[0, 5]$ & $0.815_{-0.062}^{+0.054}$ & $0.863_{-0.065}^{+0.050}$  \\\hline
\end{tabular}
\caption{The table shows the NP parameters, their prior ranges, and the constrained maximum likelihood values with uncertainty estimates at the $90\%$ credible intervals in the mass shift prescription.}
\label{tab:kt}
\end{table}

Fig.~\ref {fig:theta_Bayes} illustrates the posterior distributions of the different parameters in the $\theta$ prescription, as obtained from the Bayesian inference of the experimental data based on two independent data sets. In these figures, the diagonal panels represent the posterior probability density distributions of each parameter, while the off-diagonal panels display the correlation probability density distributions between two different parameters. Meanwhile, we present maximum a posteriori (MAP) with the $90\%$ CI of the extracted model parameters in Tab.~\ref{tab:theta}, obtained by excluding the lowest and highest $5\%$ regions of their distribution functions. The results indicate that ISR effects are negligible, with hadronization and UE contributions dominating. Furthermore, the MAP values extracted from two independent data sets agree to within small relative uncertainties, demonstrating the robustness of our Bayesian analysis.

In addition, we show the corresponding posterior distributions and the correlation probability density distributions of the parameters in the mass shift prescription in Fig.~\ref{fig:kt_Bayes}, and provide the corresponding posterior probability density distributions and MAP with $90\%$ CI for each parameter in Tab.~\ref{tab:kt}. Again, the close agreement between MAP values from the two data sets confirms the stability of our extraction.

\begin{figure}[!h]
\includegraphics[width=0.85\linewidth]{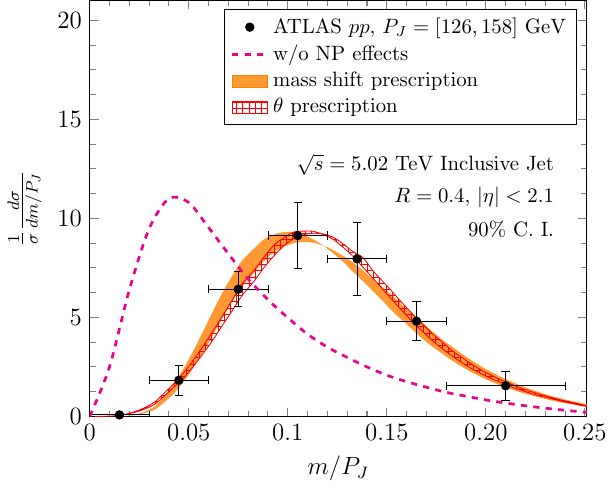}
\caption{Comparison numerical results with the ATLAS measurement of the jet mass distribution in $pp$ collisions~\cite{ATLAS:2018jsv}.}
\label{fig:atlasNP1}
\end{figure}

\begin{figure}[!h]
\includegraphics[width=0.85\linewidth]{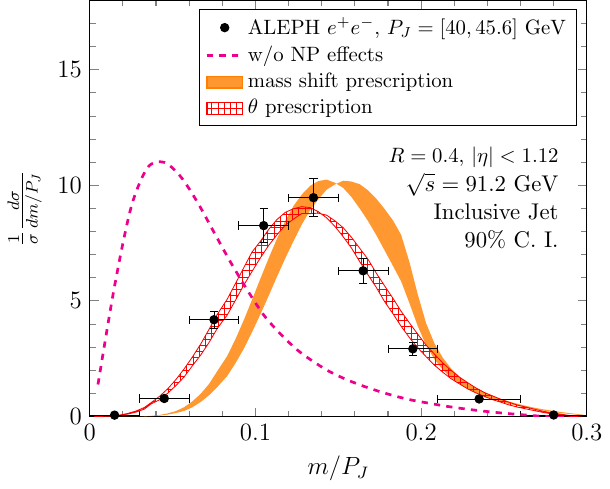}
\caption{Comparison numerical results with the ALEPH measurement of the jet mass distribution in $e^+e^-$ collisions~\cite{Chen:2021uws}.}
\label{fig:ALEPHNP}
\end{figure} 

\textit{Numerical results---} As shown in Fig.~\ref{fig:atlasNP1}, we compare the numerical results with the ATLAS measurement of the jet mass distribution in $pp$ collisions~\cite{ATLAS:2018jsv}. The magenta dashed line represents the resummation calculations without NP effects, while the red and orange bands include NP effects, corresponding to the $\theta$ and mass shift prescriptions, respectively. Without the NP contribution, the resummation formalism underestimates the jet mass data. Meanwhile, the NP contribution significantly enhances the jet mass, shifting the peak to a larger mass region. Our analysis, using the maximum of posterior distributions with a $90\%$  CI, describes the ATLAS data well.

In $e^+e^-$ annihilations, the absence of ISR and UE effects leaves hadronization as the sole NP effect; thus, the inclusive jet mass measurements in $e^+e^-$ collisions can serve as an ideal `golden channel' for validating our Bayesian analysis. After removing the ISR and UE contributions and including only the hadronization contribution determined from our Bayesian fit, we achieve excellent agreement with the ALEPH archived data for $e^+e^-$ collisions at 91.2 GeV~\cite{Chen:2021uws}, as shown in Fig.~\ref{fig:ALEPHNP}. Notably, since we did not include any $e^+e^-$ data in our fit, this agreement provides a genuine test of the predictive power of this NP model study.

Finally, we observe that the NP parameters in the momentum shift prescription, extracted via Bayesian inference, exhibit considerable sensitivity to the inclusion of CMS dijet data. When fitted without dijet data, these parameters lead to a modest overestimate of the jet mass distributions in both CMS $pp$ dijet collisions~\cite{CMS:2018vzn} and ALEPH $e^+e^-$ single-jet data~\cite{Chen:2021uws}. In contrast, the new 3D NP prescription demonstrates robustness: it yields parameters that are insensitive to whether dijet data are included and that consistently describe both CMS dijet and ALEPH measurements. A comprehensive numerical comparison is provided in the Supplemental Material.

\textit{Conclusion---} In this paper, we introduce a novel three-dimensional NP model for the jet mass observable, in addition to the conventional momentum shift NP model. By performing a Bayesian analysis of jet mass measurements in $pp$ collisions at RHIC and LHC energies, we simultaneously obtain the NP contributions from hadronization, ISR, and UE effects with two independent NP prescriptions. Our analysis provides an excellent description of all relevant $pp$ data. Furthermore, as a cross-check, when only the hadronization effect is present, we also achieve good agreement with the ALEPH inclusive-jet mass spectrum at $\sqrt{s}=91.2$ GeV in $e^+e^-$ collisions. This study may help pin down NP physics in precision QCD studies and thus could serve as a benchmark example for other jet substructure observables. Lastly, the NP model developed in this work can also contribute to the quantitative understanding of soft physics across a wide range of jet substructure measurements.

\textit{Acknowledgments---}
We thank Gregory Soyez, Wen-jing Xing, Feng Yuan and Kai Zhou for useful inputs and discussions.  This work is supported in part by the Ministry of Science and Technology of China under Grant No. 2024YFA1611004, by the CUHK Shenzhen University Development Fund under Grant No.
 UDF01001859, and by National Natural Science Foundation of China under Grants No.11935007.

\bibliographystyle{apsrev4-1}

\newpage
\begin{widetext}

\let\oldaddcontentsline\addcontentsline
\renewcommand{\addcontentsline}[3]{}
\section*{Supplemental material}

\let\addcontentsline\oldaddcontentsline

\tableofcontents

\vspace{1.cm}

In this Supplemental Material, we provide the details of the Bayesian analysis. Our theoretical framework, featuring two NP parametrizations for the jet invariant-mass spectrum, is presented in Sec.~\ref{sec::frame}. Section~\ref{sec::sets} introduces the two independent datasets, each covering a different data range for our Bayesian inference. In Sec.~\ref{sec::cms}, we discuss the impact of the CMS dijet data on the NP values extracted from our Bayesian analysis using these two parametrizations.  Finally, Sec.~\ref{sec::add} presents additional numerical results, comparing both parametrizations with the experimental measurements.

\section{The theoretical framework}
\label{sec::frame}

Within the collinear‐factorization framework, the differential jet invariant‐mass spectrum can be expressed as
\begin{eqnarray}
\frac{d\sigma}{dm_0} = \int x_a f(x_a) \otimes x_b f(x_b) \otimes \mathcal{H}(\hat s, \hat t, \hat u) \otimes \mathcal{S}(P_J, m_0, R),
\end{eqnarray}  
where $x_{a,b}$ denote the longitudinal momentum fractions of the incoming partons, $m_0$ is the partonic invariant mass, $P_J$ stands for the jet transverse momentum, and $R$ represents the jet‐cone size. The functions $x f(x)$ represent the collinear parton distribution functions (PDFs). In our calculation, we use the NLO CTEQ set~\cite{Hou:2019efy}. The hard-scattering coefficient $\mathcal{H}(\hat{s}, \hat{t}, \hat{u})$ depends on the usual Mandelstam variables ($\hat{s}$, $\hat{t}$, and $\hat{u}$), while the soft function $\mathcal{S}(P_J, m_0, R)$ encodes the Sudakov factor from final-state soft-collinear radiation. The Sudakov factor is given by  
\begin{equation}
\mathcal{S}(P_J, m_0, R)=\frac{\partial \Sigma_{\rm res}\left( {m_0^2}/({\pperp ^2 R^2 })\right) }{\partial m_0}. 
\end{equation}  
Within the eikonal approximation and including running‐coupling effects at next‐to‐leading logarithmic accuracy (NLL), one finds
\begin{equation}
\Sigma_{\rm res}(\rho )=\exp \left[ -\int ^1_{\rho} \frac{\dd \rho ^\prime}{\rho ^\prime} \int ^1_{\rho ^\prime} \dd \xi \frac{\alpha_s( \pperp R  \sqrt{\xi  \rho ^\prime})}{2\pi} P_i(\xi)\right],
\end{equation}
with  $\rho={m_0^2}/({	\pperp ^2 R^2 })$ and $P_i(\xi)$ the final‐state splitting function. To include soft two‐loop effects, we adopt the Catani–Marchesini–Webber (CMW) scheme~\cite{Catani:1990rr}, in which
\begin{equation}
\alpha_s ^{\text{CMW}}(\mu )  =\alpha_s(\mu) + \frac{K}{2\pi} \alpha_s ^2(\mu),
\end{equation}
with two-loops cusp anomalous dimension $K = C_A\left(\frac{67}{18} -\frac{\pi ^2}{6}  \right)-\frac{5}{9}n_f$. To extend these analytic results into the deep nonperturbative region, we freeze the coupling below an NP scale following Ref.~\cite{Marzani:2019hun}
\begin{equation}
\bar \alpha_s(\mu)
=
 \alpha_s(\mu) \theta (\mu - \mu_{\text{NP}})
 +
  \alpha_s(\mu_{\text{NP}}) \theta (\mu_{\text{NP}} - \mu).
\end{equation}
With this coupling prescription, the Sudakov form factor acquires a modified expression, as detailed in Refs.~\cite{Dasgupta:2015lxh,Marzani:2019hun}.

Since the above resummed result captures only perturbative soft‐gluon radiation, we incorporate  NP models to describe the leading soft‐physics corrections. In this work, we consider two NP parametrizations: the novel three-dimensional angular prescription (referred to as the $\theta$ prescription for short) developed here, and the mass shift prescription due to one-dimensional NP momentum increase (referred to as the mass shift prescription for short) given by Ref.~\cite{Stewart:2014nna}.

\begin{itemize}
\item
As presented in the main text, the jet invariant-mass spectrum in the three-dimensional $\theta$ prescription reads
\begin{eqnarray}
\frac{d\sigma}{dm} &=&
\int   d\phi d  m_{\rm NP} m_{\rm NP} \frac{d\sigma}{dm_{0}}\frac{  m }{\sqrt{m^{2}-m_{\rm NP}^{2}(1-\cos ^2\phi)} }  G(m_{\rm NP}),
\end{eqnarray}
with $f(m_0)= \sqrt{ m_0 ^{2}+2 m_0 m_{\rm NP}  \cos\phi+
m_{\rm NP}^{2}}$ and $m_{0}=- m_{\rm NP} \cos \phi \pm
\sqrt{m^{2}-m_{\rm NP}^{2}(1-\cos ^2\phi)} \geq 0 $.
We assume the non-perturbative function $G(m_{\rm NP})$ satisfies the Gaussian distribution as
\begin{equation}
G(m_{\rm NP})=\frac{1}{\pi\lambda^{2}}e^{-\frac{m_{\rm NP}^{2}}{\lambda^{2}}},  \qquad {\rm with} \qquad
\lambda^2 =Q^2_{\rm NP} \frac{P^2_J}{m^2},
\end{equation}
where $Q_{\rm NP}$ stands for the NP factor. Under the assumption that hadronization, ISR, and UE contributions factorize independently, the total NP factor is given by
\begin{equation}
Q_{\rm NP}^{2} = Q_{h}^{2} +Q_{i}^{2} + Q_{\rm UE}^{2},
\end{equation}
where $Q_{h}$ denotes the final‐state hadronization contribution, $Q_{i}$ the initial‐state radiation effect, and $Q_{\rm UE}$ the underlying‐event contribution arising from the collision background or multi‐parton interactions, which grows with the center-of-mass energy. We adopt the following ansatz for these contributions
\begin{eqnarray}
Q^2_{ i}=(a^2_{i, a}+a^2_{i, b} ) R^4\,{\rm GeV}^2, \quad
Q^2_{ h}=&a^2_{h,c} R^2 \,{\rm GeV}^2, \quad
Q^2_{\rm UE}=a^2_{\rm UE} \sqrt{s} R^4\,{\rm GeV}^2,
\end{eqnarray}
where the dimensionless NP coefficients $a_{i,a}$, $a_{i,b}$, $a_{h,c}$, and $a_{\rm UE}$ are determined by experimental data with $a, b, c=q, g$, and the $\sqrt s$ is expressed in units of TeV. For the $g+q\rightarrow g+X$ channel, we write
\begin{eqnarray}
Q_{\rm NP }^{2} \Big|_{g+q\rightarrow g+X}= a^2_{h,g} R^2 + (a^2_{i, g}+a^2_{i, q} ) R^4+ a^2_{\rm UE} \sqrt{s} R^4.
\end{eqnarray}

\item
In the mass shift prescription, the modified jet mass distribution by modelling parton-to-hadron transverse momentum transitions is given by~\cite{Stewart:2014nna}
\begin{equation}
\frac{d\sigma}{dm} = \int d k_t \frac{d\sigma}{dm_0} F_k(k_t), \quad \text{with} \quad  m^2_0=m^2-2 k_t P_J,
\end{equation}
where the shape function is given as
\begin{equation}
F_k(k_t) =\frac{4k_t }{\Omega_{\rm NP}} e^{-\frac{2k_t}{\Omega_{\rm NP}}}.
\end{equation}
The non-perturbative factor also includes the different non-perturbative effects, which is given as
\begin{equation}
\Omega_{\rm NP} = \Omega_{ h} + \Omega_{ i} + \Omega_{\rm UE},
\end{equation}
where the first term $\Omega_{ h}$,  $\Omega_{ i}$, and  $\Omega_{\rm UE}$ denote hadronization, ISR, and UE effects, respectively. The parametrization form is given as
\begin{eqnarray}
\Omega_{ i} =(b_{i, a} +b_{i,b}) R^4 \, {\rm GeV}, \quad
\Omega_{ h}=&b^{(1)}_{h,c} R+b^{(3)}_{h,c} R^3\, {\rm GeV}, \quad
\Omega_{\rm UE} =b_{\rm UE} \sqrt{s} R^4 \, {\rm GeV},
\end{eqnarray}
with  $b_{i,a}$, $b_{i,b}$, $b_{h,c}^{(1)}$, $b_{h,c}^{(3)}$, and $b_{\rm UE}$ are NP coefficients to be fitted. For the $g+q\rightarrow g+X$ channel, one finds
\begin{eqnarray}
\Omega_{\rm NP } \Big|_{g+q\rightarrow g+X}= b^{(1)}_{h,g} R + b^{(3)}_{h,g} R^3 + (b_{i, g}+b_{i, q} ) R^4+ b_{\rm UE} \sqrt{s} R^4.
\end{eqnarray}

\end{itemize}

To relate quark and gluon NP parameters due to different color factors, in the $\theta$ prescription we impose 
\begin{equation}
a_{h,q}^2=\frac{C_F}{C_A} a_{h,g}^2, \qquad a_{i,q}^2=\frac{C_F}{C_A} a_{i,g}^2,
\end{equation}
while in the mass shift prescription
\begin{equation}
b^{(1)}_{h,q}=\frac{C_F}{C_A} b^{(1)}_{h,g}, \qquad b^{(3)}_{h,q}=\frac{C_F}{C_A} b^{(3)}_{h,g}, \qquad b_{i,q}=\frac{C_F}{C_A} b _{i,g}.
\end{equation}
Finally, we assume the underlying‐event contribution to be flavor‐independent, i.e., identical for quark and gluon jets.

\section{The Bayesian inference based on two independent data sets}
\label{sec::sets}

In this section we present further details of our Bayesian analysis based on two independent small‐mass datasets.  In the low‐mass region, both nonperturbative soft physics and Sudakov resummation are important, while higher‐order perturbative corrections dominate at large masses.  Consequently, we restrict our extraction to the regime $m\ll P_JR$ and exclude data from the large mass region. To test the stability of our fits, we consider two small‐mass selections defined by different upper‐mass cuts.  Tab.~\ref{tab:supp_DataSet} summarizes the measurements included: single‐inclusive jet mass data from STAR at $\sqrt{s}=200$ GeV~\cite{STAR:2021lvw}, and from ATLAS at $\sqrt{s}=7$ TeV~\cite{ATLAS:2012am} and 5.02 TeV~\cite{ATLAS:2018jsv}; and $W/Z$-tagged exclusive‐jet masses from CMS at $\sqrt{s}=7$ TeV~\cite{CMS:2013kfv}.
In the following section, we examine how the CMS dijet jet mass measurements $\sqrt{s}=13$ TeV~\cite{CMS:2018vzn} influence the NP values extracted via Bayesian inference under the two NP prescriptions.

\begin{table}[ht]
    \centering
\begin{tabular}{|c|c|c|c|c|c|}
\hline
&\multirow{2}{*}{}                            &\multirow{2}{*}{Jet radius}   & \multirow{2}{*}{$P_J$ [GeV]}  &\multicolumn{2}{c|}{The data ranges ($m$ [GeV]) } \\ \cline{5-6} 
&                                             &                              &               & Set $1$       &  Set $2$     \\ \hline 
\multirow{4}{*}{(i)}&\multirowcell{4}{CMS \\ 7 TeV \\ Z/W+jet}  &\multirow{4}{*}{$R=0.7$}                & [125, 150]    & [0, 40]     & [0, 50]     \\ \cline{4-6} 
&                                             &                                                          & [150, 220]    & [0, 40]    & [0, 50]     \\ \cline{4-6} 
&                                             &                                                          & [220, 300]    & [0, 50]    & [0, 60]     \\ \cline{4-6}
&                                             &                                                          & [300, 450]    & [0, 70]    & [0, 80]    \\ \hline    
\multirow{4}{*}{(ii)}&\multirowcell{4}{ATLAS \\ 7 TeV \\ inclusive jet} & \multirow{4}{*}{$R=1.0$}& [200, 300]    & [10, 80]    & [10, 90]   \\ \cline{4-6} 
&                                                   &                                             & [300, 400]    & [10, 90]    & [10, 100]   \\ \cline{4-6} 
&                                                   &                                             & [400, 500]    & [20, 100]    & [20, 110]   \\ \cline{4-6}
&                                                   &                                             & [500, 600]    & [20, 110]    & [20, 120]   \\ \hline    
\multirow{5}{*}{(iii)}&\multirowcell{5}{STAR \\ 0.2 TeV \\ inclusive jet} &  $R=0.2$  & [20, 25]    & [0, 5]     & [0, 6]  \\ \cline{3-6} 
&                                                          & \multirow{3}{*}{$R=0.4$} & [20, 25]    & [0, 5]     & [0, 6]  \\ \cline{4-6} 
&                                                          &                          & [25, 30]    & [0, 6]     & [0, 7]  \\ \cline{4-6}
&                                                          &                          & [30, 40]    & [0, 7]    & [0, 8] \\ \cline{3-6}
&                                                          & $R=0.6$                  & [30, 40]    & [1, 10]    & [1, 11] \\ \hline
\multirow{8}{*}{(iv)}&\multirowcell{8}{ATLAS \\ 5.02 TeV \\ inclusive jet}  &\multirow{8}{*}{$R=0.4$}   & \multirow{2}{*}{$P_J$ [GeV]}   & \multicolumn{2}{c|}{The data ranges ($m/P_J$)}   \\ \cline{5-6} 
&                                                    &                              &               & Set $1$        &  Set $2$     \\ \cline{4-6}   
&                                                    &                              & [126, 158]    & [0, 0.18]    & [0, 0.21]   \\ \cline{4-6} 
&                                                    &                              & [158, 199]    & [0, 0.18]    & [0, 0.21]   \\ \cline{4-6} 
&                                                    &                              & [199, 251]    & [0, 0.15]    & [0, 0.18]   \\ \cline{4-6}
&                                                    &                              & [251, 316]    & [0, 0.15]    & [0, 0.18]   \\ \cline{4-6} 
&                                                    &                              & [316, 398]    & [0, 0.15]    & [0, 0.18]   \\ \cline{4-6}
&                                                    &                              & [398, 500]    & [0, 0.15]    & [0, 0.18]   \\ \hline    
\end{tabular}
\caption{Ranges of the two experimental datasets used in the Bayesian analysis. The measurements include: (i) CMS data at $\sqrt{s} = 7~\text{TeV}$ for the W/Z-tagged exclusive jet mass~\cite{CMS:2013kfv}, (ii) ATLAS data at $\sqrt{s} = 7~\text{TeV}$ \cite{ATLAS:2012am}, (iii) STAR data at $\sqrt{s} = 200~\text{GeV}$ \cite{STAR:2021lvw}, (iv) ATLAS data at $\sqrt{s} = 5.02~\text{TeV}$ for the inclusive jet mass measurement~\cite{ATLAS:2018jsv}.}
\label{tab:supp_DataSet}
\end{table}

\begin{table}[!ht]
\begin{tabular}{|c|c|c|c| }
\hline
\multirow{2}{*}{Parameter}  & \multirow{2}{*}{Prior range} & \multicolumn{2}{c|}{MAP}  \\\cline{3-4} 
          &              & Set $1$         & Set $2$         \\\hline
$a_{h,g}$      & $[0, 5]$ & $2.907_{-0.169}^{+0.127}$ & $2.637_{-0.178}^{+0.113}$   \\\hline
$a_{i,g}$                & $[0, 5]$ & $0.180_{-0.146}^{+0.745}$ & $0.324_{-0.259}^{+0.911}$   \\\hline
$a_{h,q}$      & $[0, 5]$ & $1.938_{-0.112}^{+0.085}$ & $1.758_{-0.119}^{+0.075}$  \\\hline
$a_{i,q}$                  & $[0, 5]$ & $0.120_{-0.097}^{+0.497}$ & $0.216_{-0.172}^{+0.607}$  \\\hline
$a_{\rm UE}$          & $[0, 5]$ & $3.712_{-0.107}^{+0.129}$ & $3.856_{-0.109}^{+0.108}$  \\\hline
\end{tabular}
\caption{
Nonperturbative parameters in the $\theta$ prescription: prior ranges and maximum‐likelihood estimates with $90\%$ CI uncertainties. }
\label{tabap:theta}
\end{table}

\begin{table}[!ht]
\begin{tabular}{|c|c|c|c|  }
\hline
\multirow{2}{*}{Parameter} & \multirow{2}{*}{Prior range} & \multicolumn{2}{c|}{MAP}  \\\cline{3-4} 
                         &          & Set $1$   & Set $2$ \\\hline
$b_{h,g}^{(1)}$    & $[0, 5]$ & $1.530_{-0.116}^{+0.086}$ & $1.494_{-0.125}^{+0.080}$ \\\hline
$b_{h,g}^{(3)}$    & $[0, 5]$ & $0.084_{-0.055}^{+0.829}$ & $0.224_{-0.185}^{+0.800}$ \\\hline
$b_{i,g}$         & $[0, 5]$ & $0.018_{-0.013}^{+0.222}$ & $0.018_{-0.013}^{+0.206}$ \\\hline
$b_{h,q}^{(1)}$    & $[0, 5]$ & $0.680_{-0.052}^{+0.038}$ & $0.664_{-0.055}^{+0.035}$ \\\hline
$b_{h,q}^{(3)}$  & $[0, 5]$ & $0.037_{-0.024}^{+0.368}$ & $0.100_{-0.082}^{+0.356}$ \\\hline
$b_{i,q}$         & $[0, 5]$ & $0.008_{-0.006}^{+0.099}$ & $0.008_{-0.006}^{+0.092}$ \\\hline
$b_{\rm UE}$     & $[0, 5]$ & $0.815_{-0.063}^{+0.055}$ & $0.866_{-0.068}^{+0.047}$  \\\hline
\end{tabular}
\caption{Nonperturbative parameters in the mass shift prescription: prior ranges and maximum‐likelihood estimates with $90\%$ CI uncertainties. }
\label{tabap:kt}
\end{table}

\begin{figure}[!h]
\includegraphics[width=0.4\linewidth]{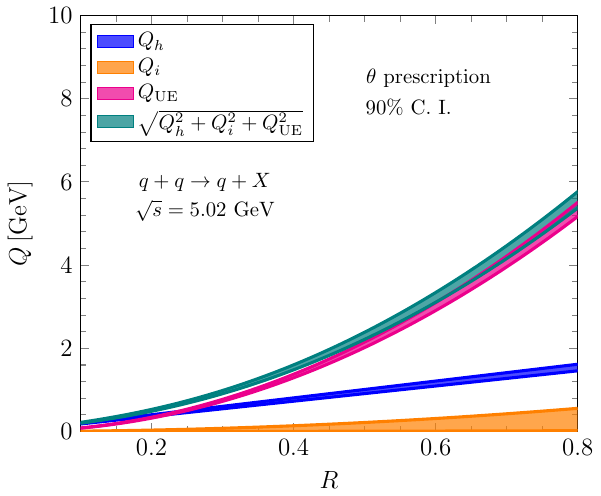}
\includegraphics[width=0.4\linewidth]{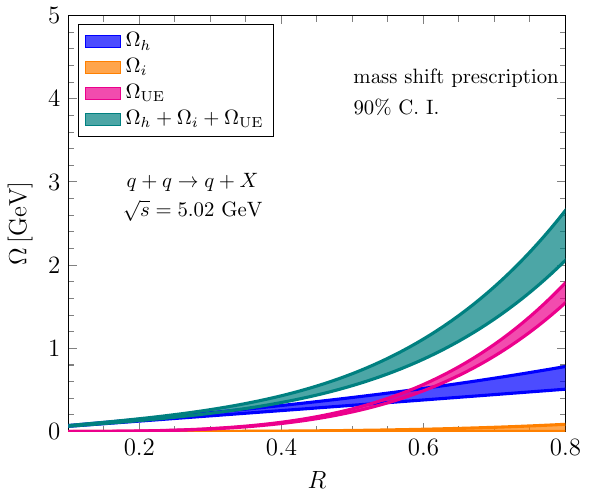}
\caption{The plots of the NP effects as functions of jet cone size $R$ at $\sqrt{s}=5.02$ TeV for the $q+q\rightarrow q+ X$ channel. The left plot is based on the $\theta$ prescription, and the right plot is based on the mass shift prescription. }
\label{fig:Que}
\end{figure}

Tab.~\ref{tabap:theta} lists the MAP estimates and their $90\%$ credible intervals (CI) for NP parameters extracted under $\theta$ prescription, while Tab.~\ref{tabap:kt} presents the corresponding results for the mass shift prescription. In both  NP parametrizations,  independent Bayesian analyses are performed on two datasets defined by different upper-mass cuts ( Tab.~\ref{tab:supp_DataSet}).  The MAP estimates from the two fits agree within small relative uncertainties in both prescriptions, confirming the stability of our extraction.

In Fig.~\ref{fig:Que}, based on the results from our Bayesian analysis, we show the NP contributions as functions of the jet cone size $R$ at $\sqrt{s}=5.02$ TeV
for the $q+q \rightarrow q+X$ channel.  The left (right) panel corresponds to the three-dimensional $\theta$ prescription (mass shift prescription). For both prescriptions, the bands correspond to the MAP values and their $90\%$ credible CI of NP parameters. In both prescriptions, the ISR contribution is found to be negligible.  In the small-$R$ regime, hadronization effects dominate while UE contributions remain minimal; as $R$ increases, the UE term becomes progressively more significant.  We emphasize that the UE parameters extracted here apply only to inclusive/exclusive jet-mass measurements in $pp$ collisions, and different collision systems ($pA$, $AA$) will require their own UE parametrizations.

\section{The impact of the CMS data from dijet events in Bayesian inference }
\label{sec::cms}

\begin{table}[ht]
    \centering
\begin{tabular}{|c|c|c|c|}
\hline
\multirowcell{13}{CMS \\ 13 TeV \\ dijet} & Jet radius                 & $P_J$ [GeV] & The data ranges ($m$ [GeV])  \\ \cline{2-4}      
                                          & \multirow{12}{*}{$R=0.8$}  & [200, 260]        & [20, 80]   \\ \cline{3-4} 
                                          &                            & [260, 350]        & [20, 80]   \\ \cline{3-4} 
                                          &                            & [350, 460]        & [20, 80]   \\ \cline{3-4}
                                          &                            & [460, 550]        & [20, 100]  \\ \cline{3-4} 
                                          &                            & [550, 650]        & [20, 100]  \\ \cline{3-4} 
                                          &                            & [650, 760]        & [40, 150]  \\ \cline{3-4} 
                                          &                            & [760, 900]        & [40, 150]  \\ \cline{3-4}
                                          &                            & [900, 1000]       & [40, 150]  \\ \cline{3-4} 
                                          &                            & [1000, 1100]      & [40, 200]  \\ \cline{3-4} 
                                          &                            & [1100, 1200]      & [40, 200]  \\ \cline{3-4} 
                                          &                            & [1200, 1300]      & [40, 200]  \\ \cline{3-4} 
                                          &                            & [1300, $+\infty$) & [40, 200]  \\ \hline
\end{tabular}
\caption{ The ranges of CMS data used in Bayesian analysis at $\sqrt{s} = 13~\text{TeV}$ for the jet mass measurements in dijet events~\cite{CMS:2018vzn}. }
\label{tabap:Dijetdata}
\end{table}

\begin{figure}[!h]
\includegraphics[width=0.49\linewidth]{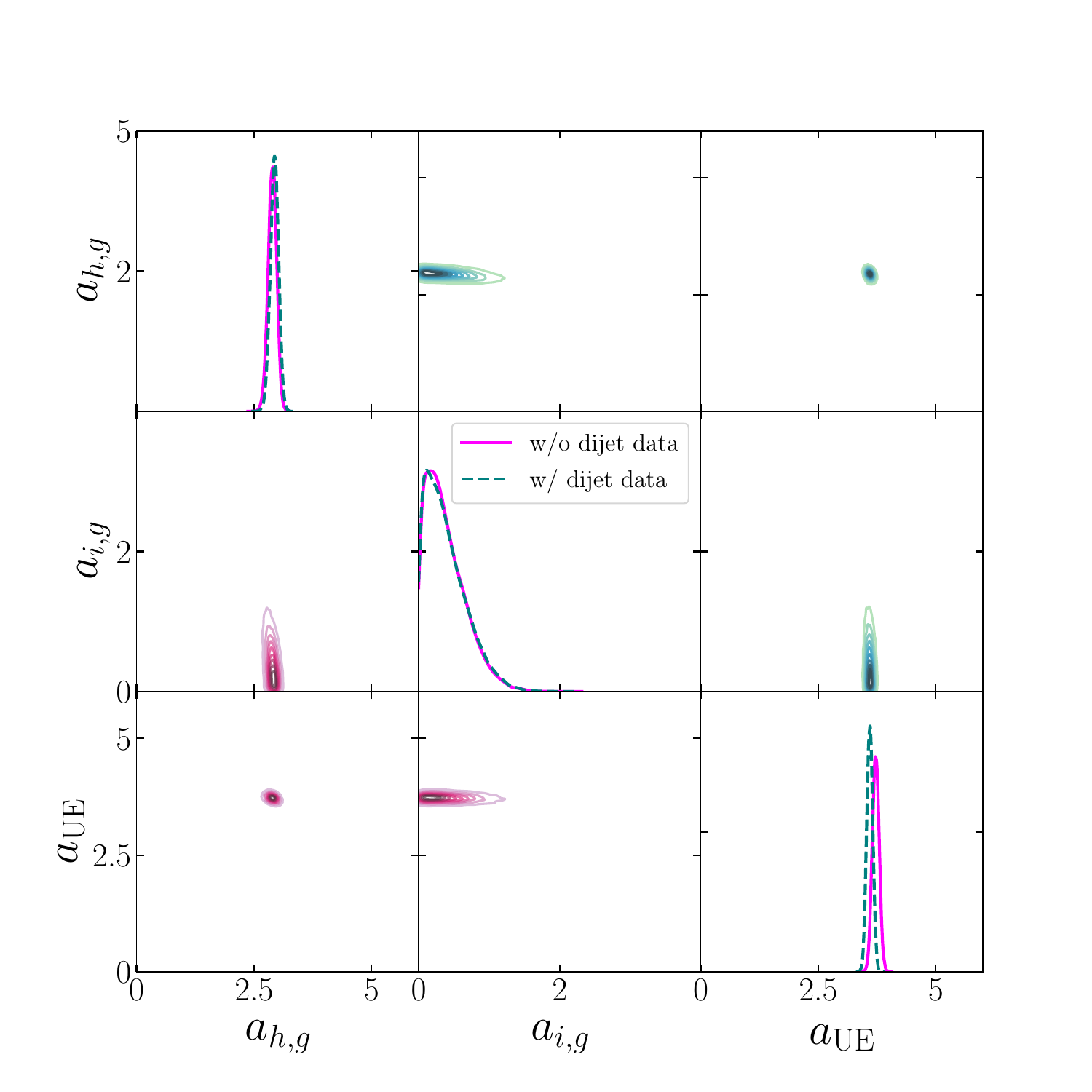}
\includegraphics[width=0.49\linewidth]{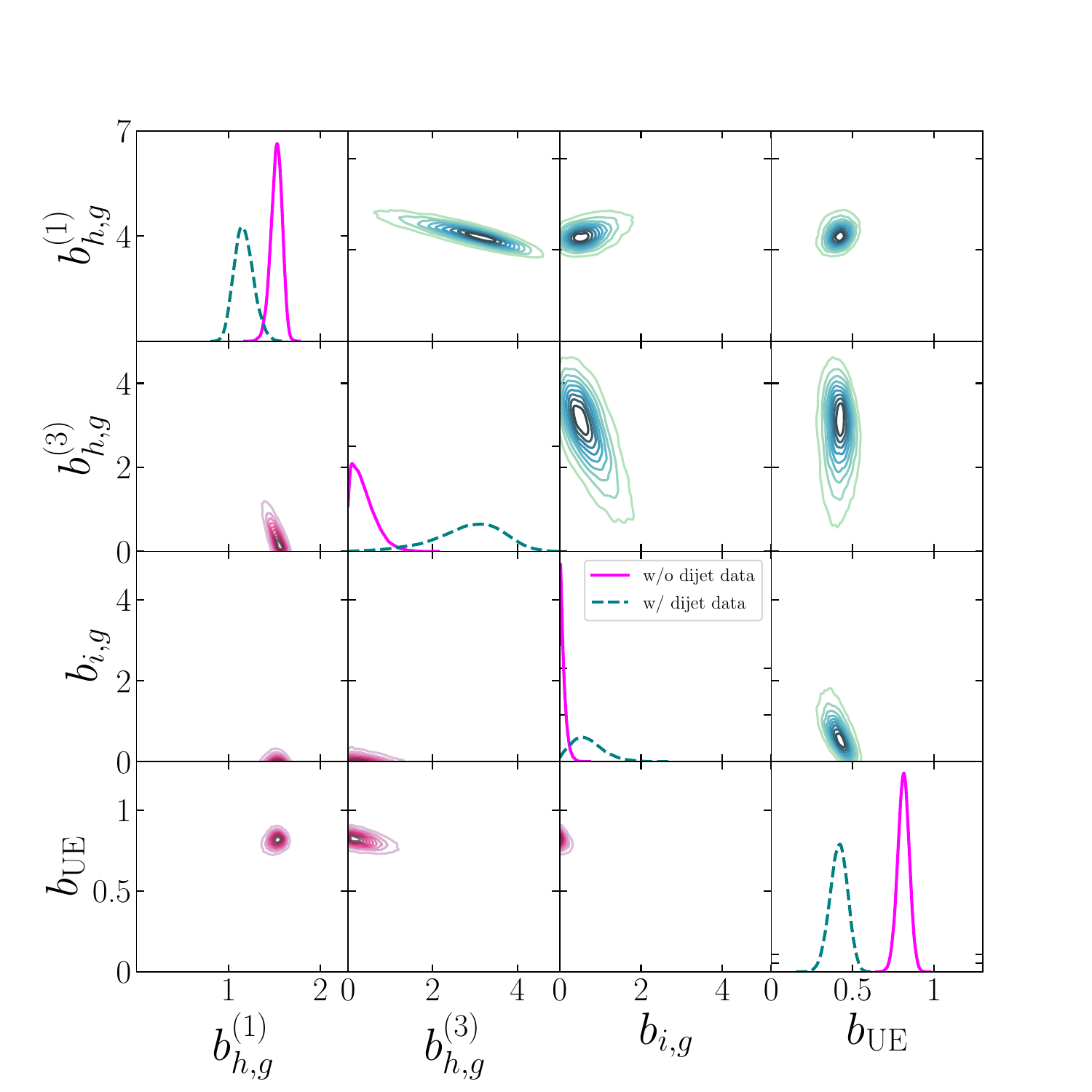}
\caption{Posterior distributions and correlations of the gluon parameters from Bayesian analysis to experimental data fitting with and without the CMS dijet data~\cite{CMS:2018vzn}. Both prescriptions used set 1 as input experimental data. Left: for $\theta$ prescription. Right: for mass shift prescription. }
\label{figap:BayesNoDijet}
\end{figure}

\begin{table}[!ht]
\begin{tabular}{|c|c|c|c|  }
\hline
\multirow{3}{*}{Parameter} & \multirow{3}{*}{Prior range} & \multicolumn{2}{c|}{MAP}  \\\cline{3-4} 
                           &                                                           & \multicolumn{2}{c|}{Data set $1$}   \\\cline{3-4} 
                                                        &                              & w/o dijet data            & w/ dijet data  \\\hline
$a_{h,g}$      & $[0, 5]$ & $2.907_{-0.169}^{+0.127}$ & $2.931_{-0.144}^{+0.145}$   \\\hline
$a_{i,g}$                  & $[0, 5]$ & $0.180_{-0.146}^{+0.745}$ & $0.104_{-0.071}^{+0.842}$   \\\hline
$a_{h,q}$      & $[0, 5]$ & $1.938_{-0.112}^{+0.085}$ & $1.954_{-0.096}^{+0.096}$  \\\hline
$a_{i,q}$                  & $[0, 5]$ & $0.120_{-0.097}^{+0.497}$ & $0.069_{-0.047}^{+0.561}$  \\\hline
$a_{\rm UE}$    & $[0, 5]$ & $3.712_{-0.107}^{+0.129}$ & $3.604_{-0.112}^{+0.096}$  \\\hline
\end{tabular}
\caption{The table displays the non-perturbative parameters of the $\theta$ prescription from Bayesian fits with and without the CMS dijet data~\cite{CMS:2018vzn}, including their prior ranges and constrained maximum likelihood values with uncertainty estimates at $90\%$ credible intervals.
}
\label{tabap:thetadijet}
\end{table}

\begin{table}[!ht]
\begin{tabular}{|c|c|c|c|c|  }
\hline
\multirow{3}{*}{Parameter}  & \multirow{3}{*}{Prior range} & \multicolumn{2}{c|}{MAP}  \\\cline{3-4} 
                           &                                                            & \multicolumn{2}{c|}{Data set $1$}   \\\cline{3-4} 
                           &                                                         & w/o dijet data            & w/ dijet data  \\\hline
$b_{h,g}^{(1)}$   & $[0, 5]$ & $1.530_{-0.116}^{+0.086}$ & $1.144_{-0.137}^{+0.206}$ \\\hline
$b_{h,g}^{(3)}$  & $[0, 5]$ & $0.084_{-0.055}^{+0.829}$ & $3.153_{-1.802}^{+0.852}$ \\\hline
$b_{i,g}$             & $[0, 5]$ & $0.018_{-0.013}^{+0.222}$ & $0.567_{-0.434}^{+0.850}$ \\\hline
$b_{h,q}^{(1)}$   & $[0, 5]$ & $0.680_{-0.052}^{+0.038}$ & $0.509_{-0.061}^{+0.092}$ \\\hline
$b_{h,q}^{(3)}$   & $[0, 5]$ & $0.037_{-0.024}^{+0.368}$ & $1.401_{-0.801}^{+0.379}$ \\\hline
$b_{i,q}$              & $[0, 5]$ & $0.008_{-0.006}^{+0.099}$ & $0.252_{-0.193}^{+0.378}$ \\\hline
$b_{\rm UE}$       & $[0, 5]$ & $0.815_{-0.063}^{+0.055}$ & $0.423_{-0.099}^{+0.078}$  \\\hline
\end{tabular}
\caption{The table displays the non-perturbative parameters of the mass shift prescription from Bayesian fits with and without the CMS dijet data~\cite{CMS:2018vzn}, including their prior ranges and constrained maximum likelihood values with uncertainty estimates at $90\%$ credible intervals.
}
\label{tabap:ktdijet}
\end{table}

\begin{figure}[!h]
\includegraphics[width=0.49\linewidth]{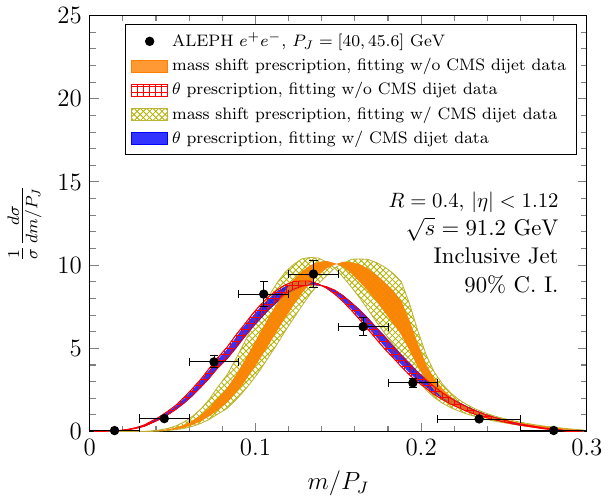}
\includegraphics[width=0.49\linewidth]{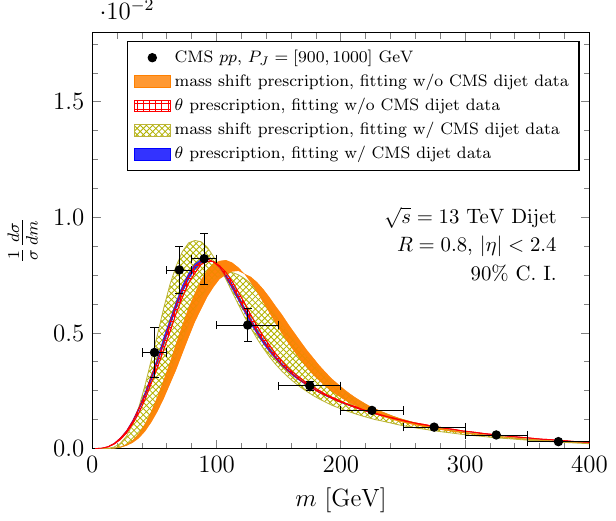}
\caption{Comparison theory results using the parameters fitting with and without the CMS dijet data~\cite{CMS:2018vzn}.  Left: for ALEPH data in $e^+e^-$ collisions at $\sqrt{s}=91.2$ GeV~\cite{Chen:2021uws}. Right: for CMS dijet data in $pp$ collisions at $\sqrt{s}=13$ TeV~\cite{CMS:2018vzn}.}
\label{figap:alephandcms}
\end{figure} 

In this section, we examine the impact of the CMS dijet data on the NP values extracted via Bayesian inference. The CMS measurement of dijet events~\cite{CMS:2018vzn} requires a dijet system with at least two jets whose leading and subleading transverse momenta, $p_{1\perp}$ and $p_{2\perp}$, satisfy: $(p_{1\perp}-p_{2\perp})/(p_{1\perp}+p_{2\perp})<0.3$ and $\Delta \phi>\pi/2$  with their azimuthal angle difference $\Delta \phi$.  

Our framework, which combines the leading-order partonic cross section with Sudakov resummation, does not explicitly implement the dijet kinematic selections. However, since the jet invariant-mass observable is governed primarily by each jet's transverse momentum and internal structure, we expect our resummed prediction to describe the CMS dijet jet-mass measurements. For both the $\theta$ and mass‐shift prescriptions, we perform two independent Bayesian analyses: one using only Data Set I (Tab.~\ref{tab:supp_DataSet}), and one including the CMS dijet measurements (Tab.~\ref{tabap:Dijetdata}).

\begin{itemize}

\item As shown in the left panel of Fig.~\ref{figap:BayesNoDijet} and summarized in Table~\ref{tabap:thetadijet}, the NP parameters extracted under the $\theta$ prescription remain essentially unchanged whether or not the CMS dijet data are included.  In Fig.~\ref{figap:alephandcms}, the red band corresponds to the extraction without CMS dijet data, while the blue band includes it; similarly, the orange band shows the resummed prediction using parameters fitted without the CMS dijet sample, and the teal band uses parameters fitted with it.  In both cases, the fitted NP parameters simultaneously describe the CMS dijet measurements in $pp$ collisions and the ALEPH single‐jet data from $e^+e^-$ collisions, demonstrating the stability and universality of the $\theta$ prescription.

\item  As shown in the right panel of Fig.~\ref{figap:BayesNoDijet} and Table~\ref{tabap:ktdijet}, the NP parameters in the mass‐shift prescription exhibit significant sensitivity to the inclusion of dijet data. Incorporating the CMS dijet measurements increases $b_{h}^{(3)}$ and $b_{i}$ while reducing $b_{h}^{(1)}$ and $b_{\rm UE}$. In Fig.~\ref{figap:alephandcms}, the orange band shows predictions using parameters fitted without the CMS dijet sample, whereas the light‐yellow band corresponds to fits that include it. The fit excluding dijet inputs modestly overestimates the jet‐mass distributions in both CMS $pp$ dijet events and ALEPH $e^+e^-$ data. By contrast, including the dijet data yields an excellent description of the CMS measurements, though it still slightly overestimates the ALEPH results.

\end{itemize}

These findings highlight the robustness of the $\theta$ prescription to variations in the CMS dijet jet-mass input, while showing that the mass-shift prescription exhibits a more noticeable response to such variations.

\section{Additional numerical results}
\label{sec::add}

In this section, we present additional numerical results compared to the experimental data. We use the NP parameters obtained from the Bayesian fit to Data Set $1$ (see Tables~\ref{tabap:theta} and \ref{tabap:kt}), which exclude the CMS dijet measurements.

 Fig.~\ref{fig:cmsZW} compares our predictions with the CMS measurement of the jet mass distribution in $Z/W+$jet events at $\sqrt{s}=7$ TeV and $R=0.7$~\cite{CMS:2013kfv}, while Fig.~\ref{fig:star} shows the comparison with STAR data in inclusive‐jet events at $\sqrt{s}=0.2$ TeV~\cite{STAR:2021lvw}. Fig.~\ref{fig:atlas5} presents the ATLAS inclusive‐jet measurement at $\sqrt{s}=5.02$ TeV with $R=0.4$~\cite{ATLAS:2018jsv}.    The red band stands for the $\theta$ prescription and the orange band for the mass shift prescription, with bands widths reflecting the MAP estimates and their $90\%$ CI from Tab.~\ref{tabap:theta} and Tab.~\ref{tabap:kt},  respectively, extracted using data set $1$. In all cases, the resummed calculations, including NP corrections under both the $\theta$ and mass shift prescriptions, provide excellent descriptions of the data.
 
Fig.~\ref{fig:atlas7} displays the ATLAS data in inclusive‐jet events at $\sqrt{s}=7$ TeV with $R=1$~\cite{ATLAS:2012am}.  As shown in Fig.~\ref{fig:cms13}, we compare our numerical predictions with the CMS measurement of the jet mass distribution in di-jet events at $\sqrt{s}=13$ TeV and $R=0.8$  in $pp$ collisions~\cite{CMS:2018vzn}. In principle, our calculations are valid only in the small-$R$ regime and are not intended to describe the jet mass in the large-$R$ region. Nevertheless, if we try to push the envelope and compare the calculation with the large-$R$ jet mass data, it is interesting to note that the $\theta$ prescription continues to agree remarkably well with the measurements, as shown in these figures.

\begin{figure*}[!ht]
\includegraphics[width=0.99\linewidth]{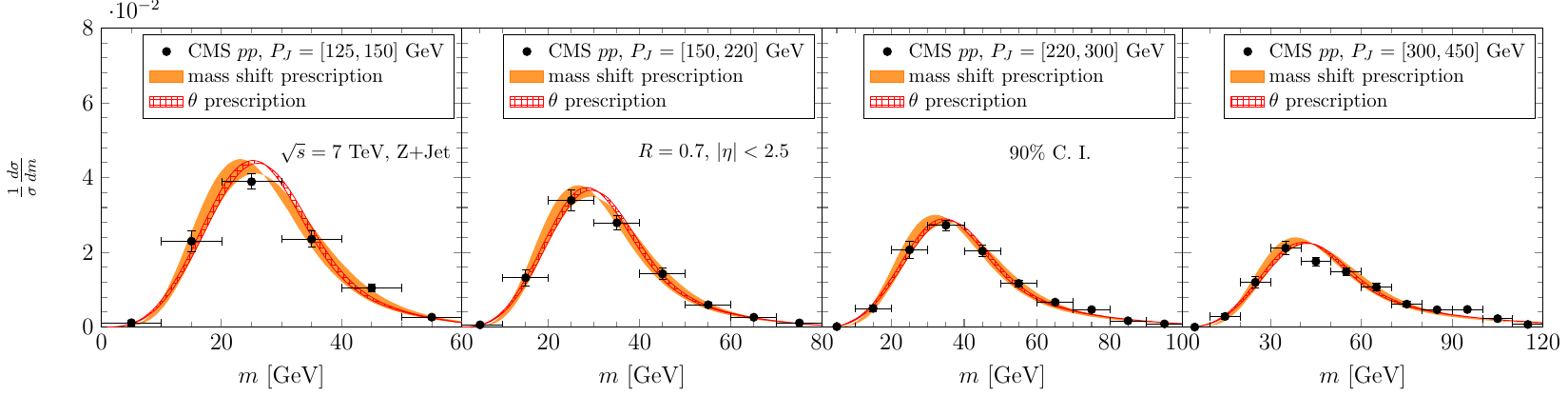}\\
\includegraphics[width=0.99\linewidth]{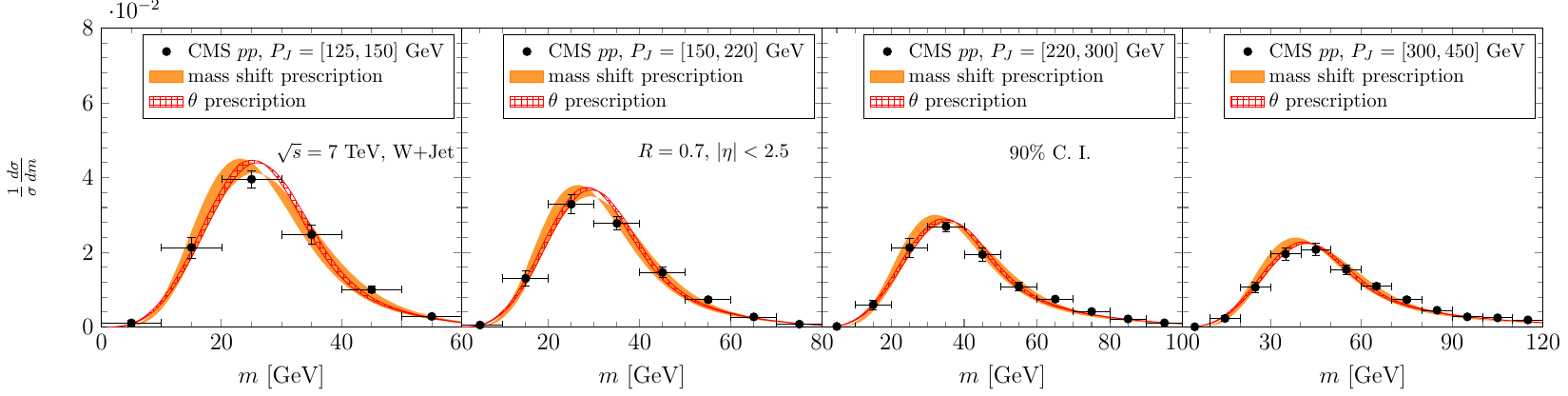}
\caption{The comparison of the numerical results with the CMS measurement of the jet mass distribution for Z/W+jet events in $pp$ collisions at $\sqrt{s}=7$ TeV and $R=0.7$~\cite{CMS:2013kfv}.}
\label{fig:cmsZW}
\end{figure*}

\begin{figure*}[!ht]
\includegraphics[width=0.75\linewidth]{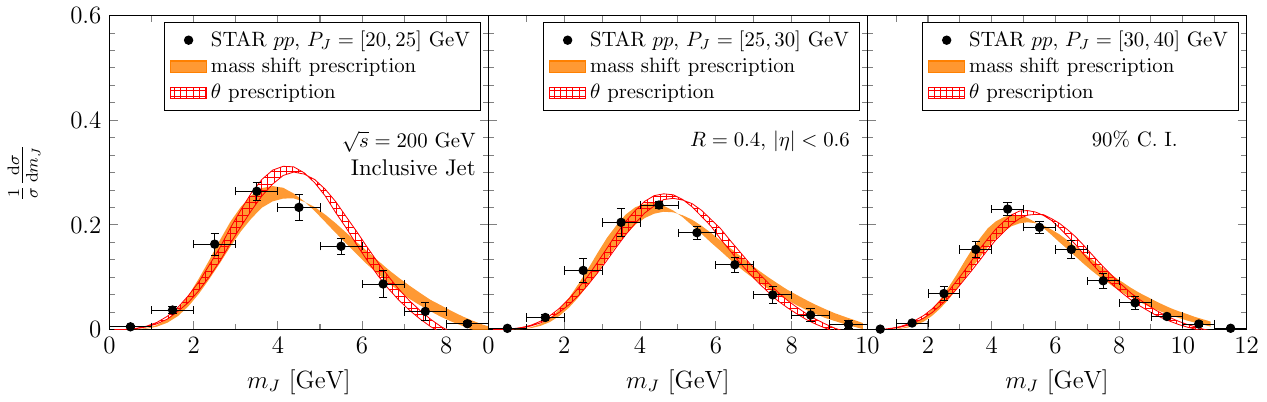}\\
\includegraphics[width=0.75\linewidth]{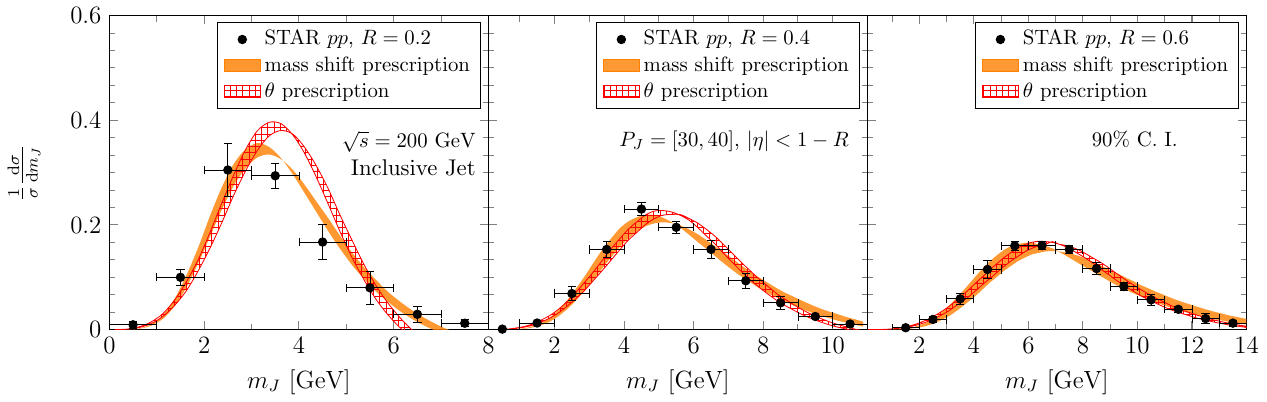}
\caption{Jet‐mass distribution in inclusive‐jet events in $pp$ collisions at $\sqrt{s}=0.2$ TeV  compared with STAR measurements~\cite{STAR:2021lvw}.}
\label{fig:star}
\end{figure*}

\begin{figure*}[!ht]
\includegraphics[width=0.75\linewidth]{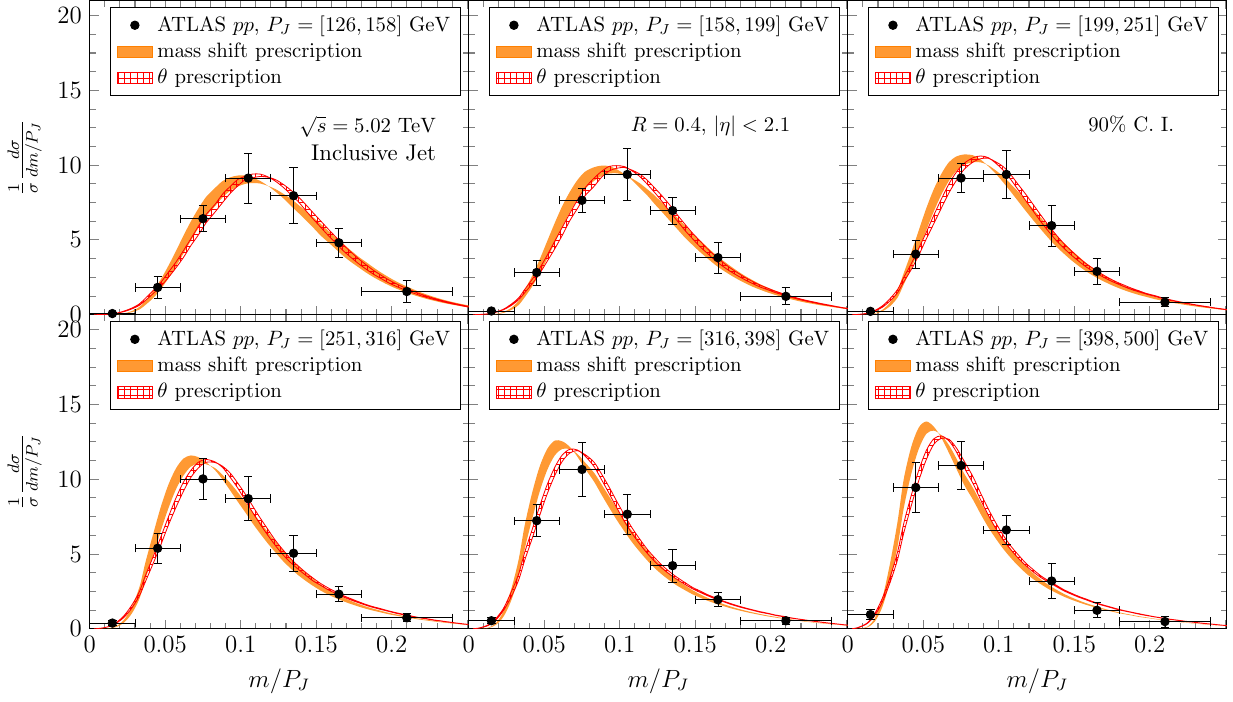}
\caption{We compare the numerical results with the ATLAS measurement of the jet mass distribution for inclusive jet events at in $pp$ collisions at $\sqrt{s}=5.02$ TeV and $R=0.4$ in $pp$ collisions~\cite{ATLAS:2018jsv}.}
\label{fig:atlas5}
\end{figure*}

\begin{figure*}[!ht]
\includegraphics[width=0.99\linewidth]{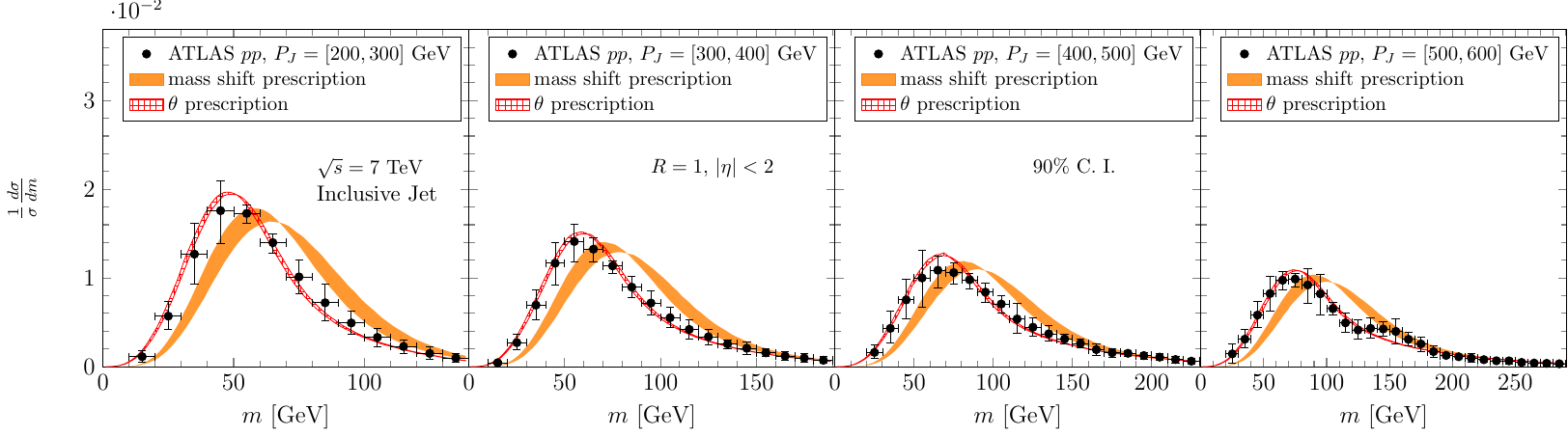}
\caption{Comparison of the numerical results with the ATLAS measurement of the jet mass distribution for inclusive jet events at in $pp$ collisions at $\sqrt{s}=7$ TeV and $R=1$ in $pp$ collisions~\cite{ATLAS:2012am}.}
\label{fig:atlas7}
\end{figure*}

\begin{figure*}[!ht]
\includegraphics[width=0.99\linewidth]{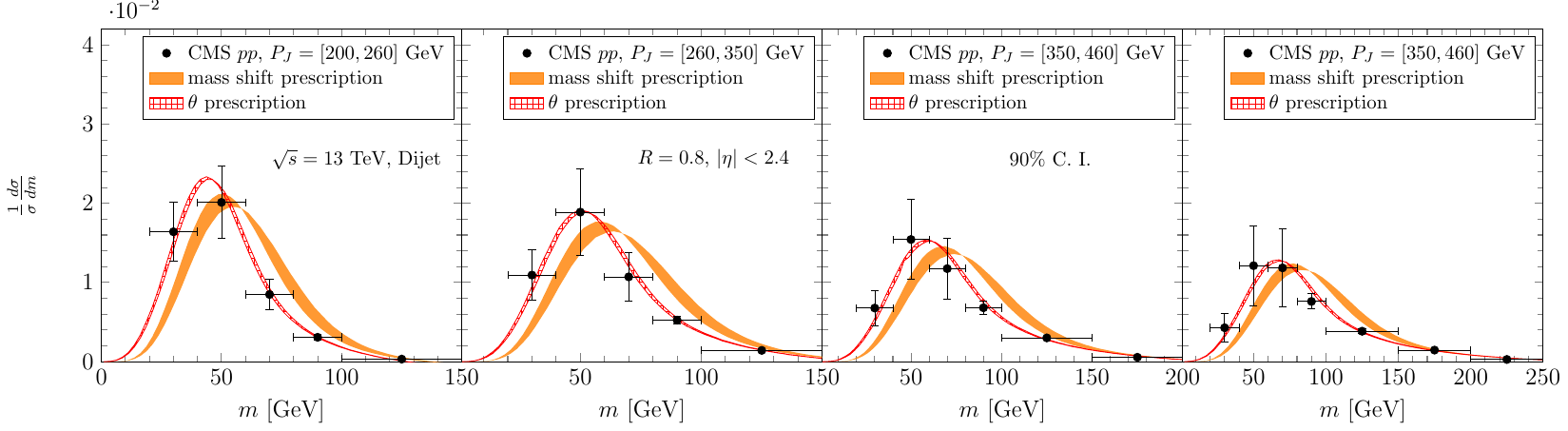}\\
\includegraphics[width=0.99\linewidth]{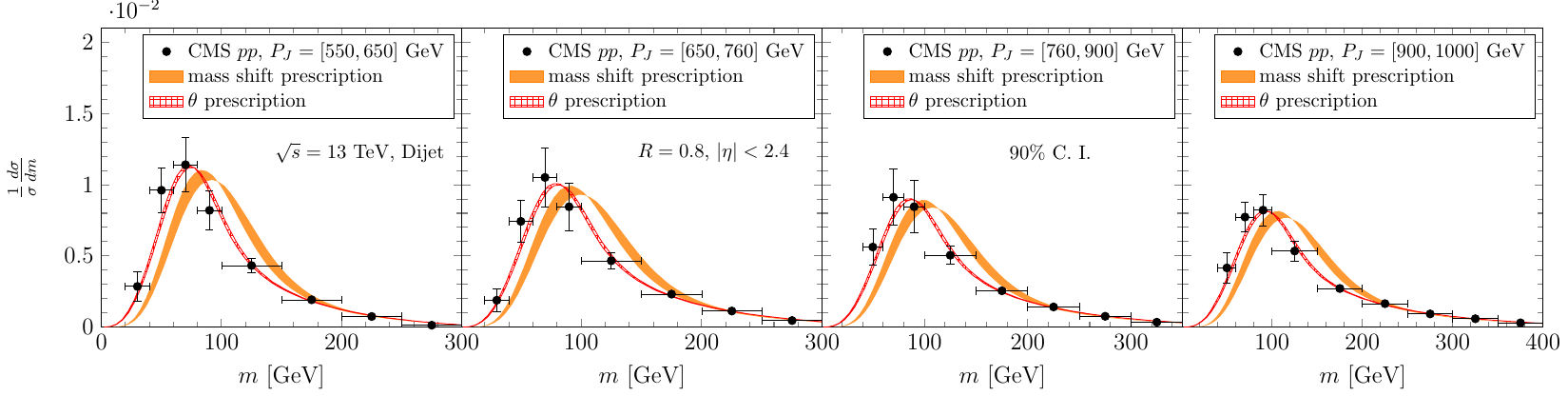}\\
\includegraphics[width=0.99\linewidth]{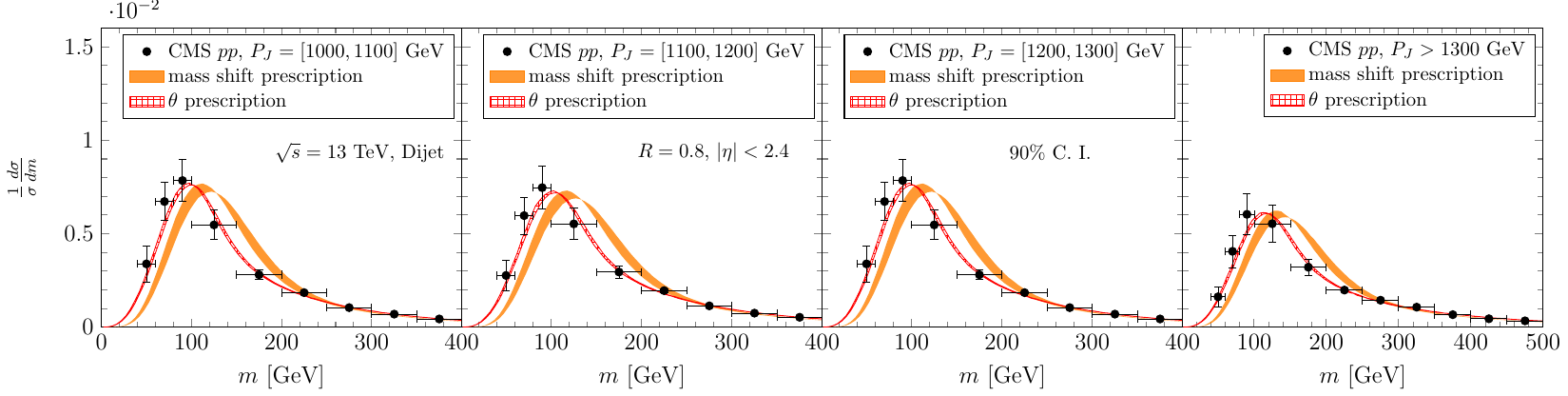}
\caption{Comparison of the numerical results with the CMS measurement of the jet mass distribution for di-jet events in $pp$ collisions at $\sqrt{s}=13$ TeV and $R=0.8$~\cite{CMS:2018vzn}.
}
\label{fig:cms13}
\end{figure*}

\end{widetext}


\begin{thebibliography}{199}
\bibitem{Pumplin:2009nk}
J.~Pumplin, J.~Huston, H.~L.~Lai, P.~M.~Nadolsky, W.~K.~Tung and C.~P.~Yuan,
Phys. Rev. D \textbf{80}, 014019 (2009)
[arXiv:0904.2424 [hep-ph]].

\bibitem{Watt:2013oha}
B.~J.~A.~Watt, P.~Motylinski and R.~S.~Thorne,
Eur. Phys. J. C \textbf{74}, 2934 (2014)
[arXiv:1311.5703 [hep-ph]].

\bibitem{ATLAS:2013pbc}
G.~Aad \textit{et al.} [ATLAS],
Eur. Phys. J. C \textbf{73}, no.8, 2509 (2013)
[arXiv:1304.4739 [hep-ex]].




\bibitem{CMS:2014qtp}
V.~Khachatryan \textit{et al.} [CMS],
Eur. Phys. J. C \textbf{75}, no.6, 288 (2015)
[arXiv:1410.6765 [hep-ex]].

\bibitem{CMS:2016lna}
V.~Khachatryan \textit{et al.} [CMS],
JHEP \textbf{03}, 156 (2017)
[arXiv:1609.05331 [hep-ex]].

\bibitem{Currie:2016bfm}
J.~Currie, E.~W.~N.~Glover and J.~Pires,
Phys. Rev. Lett. \textbf{118}, no.7, 072002 (2017)
[arXiv:1611.01460 [hep-ph]].



\bibitem{Harland-Lang:2017ytb}
L.~A.~Harland-Lang, A.~D.~Martin and R.~S.~Thorne,
Eur. Phys. J. C \textbf{78}, no.3, 248 (2018)
[arXiv:1711.05757 [hep-ph]].

\bibitem{Currie:2017eqf}
J.~Currie, A.~Gehrmann-De Ridder, T.~Gehrmann, E.~W.~N.~Glover, A.~Huss and J.~Pires,
Phys. Rev. Lett. \textbf{119}, no.15, 152001 (2017)
[arXiv:1705.10271 [hep-ph]].

\bibitem{Marzani:2019evv}
S.~Marzani, D.~Reichelt, S.~Schumann, G.~Soyez and V.~Theeuwes,
JHEP \textbf{11}, 179 (2019)
[arXiv:1906.10504 [hep-ph]].

\bibitem{Gutierrez-Reyes:2019msa}
D.~Gutierrez-Reyes, Y.~Makris, V.~Vaidya, I.~Scimemi and L.~Zoppi,
JHEP \textbf{08}, 161 (2019)
[arXiv:1907.05896 [hep-ph]].

\bibitem{AbdulKhalek:2020jut}
R.~Abdul Khalek, S.~Forte, T.~Gehrmann, A.~Gehrmann-De Ridder, T.~Giani, N.~Glover, A.~Huss, E.~R.~Nocera, J.~Pires and J.~Rojo, \textit{et al.}
Eur. Phys. J. C \textbf{80}, no.8, 797 (2020)
[arXiv:2005.11327 [hep-ph]].

\bibitem{ATLAS:2021qnl}
G.~Aad \textit{et al.} [ATLAS],
JHEP \textbf{07}, 223 (2021)
[arXiv:2101.05095 [hep-ex]].

\bibitem{CMS:2021iwu}
A.~Tumasyan \textit{et al.} [CMS],
JHEP \textbf{01}, 188 (2022)
[arXiv:2109.03340 [hep-ex]].

\bibitem{ALICE:2021njq}
S.~Acharya \textit{et al.} [ALICE],
JHEP \textbf{05}, 061 (2022)
[arXiv:2107.11303 [nucl-ex]].

\bibitem{Benitez:2024nav}
M.~A.~Benitez, A.~H.~Hoang, V.~Mateu, I.~W.~Stewart and G.~Vita,
[arXiv:2412.15164 [hep-ph]].

\bibitem{CMS:2013vbb}
S.~Chatrchyan \textit{et al.} [CMS],
Eur. Phys. J. C \textbf{73}, no.10, 2604 (2013)
[arXiv:1304.7498 [hep-ex]].

\bibitem{CMS:2014mna}
V.~Khachatryan \textit{et al.} [CMS],
Eur. Phys. J. C \textbf{75}, no.5, 186 (2015)
[arXiv:1412.1633 [hep-ex]].

\bibitem{ATLAS:2015yaa}
G.~Aad \textit{et al.} [ATLAS],
Phys. Lett. B \textbf{750}, 427-447 (2015)
[arXiv:1508.01579 [hep-ex]].

\bibitem{ATLAS:2017qir}
M.~Aaboud \textit{et al.} [ATLAS],
Eur. Phys. J. C \textbf{77}, no.12, 872 (2017)
[arXiv:1707.02562 [hep-ex]].

\bibitem{Britzger:2017maj}
D.~Britzger, K.~Rabbertz, D.~Savoiu, G.~Sieber and M.~Wobisch,
Eur. Phys. J. C \textbf{79}, no.1, 68 (2019)
[arXiv:1712.00480 [hep-ph]].

\bibitem{Hannesdottir:2022rsl}
H.~S.~Hannesdottir, A.~Pathak, M.~D.~Schwartz and I.~W.~Stewart,
JHEP \textbf{04}, 087 (2023)
[arXiv:2210.04901 [hep-ph]].

\bibitem{Benitez:2025vsp}
M.~A.~Benitez, A.~Bhattacharya, A.~H.~Hoang, V.~Mateu, M.~D.~Schwartz, I.~W.~Stewart and X.~Zhang,
[arXiv:2502.12253 [hep-ph]].

\bibitem{Soper:2010xk}
D.~E.~Soper and M.~Spannowsky,
JHEP \textbf{08}, 029 (2010)
[arXiv:1005.0417 [hep-ph]].

\bibitem{Godbole:2014cfa}
R.~M.~Godbole, D.~J.~Miller, K.~A.~Mohan and C.~D.~White,
JHEP \textbf{04}, 103 (2015)
[arXiv:1409.5449 [hep-ph]].

\bibitem{Chen:2014dma}
N.~Chen, J.~Li, Y.~Liu and Z.~Liu,
Phys. Rev. D \textbf{91}, no.7, 075002 (2015)
[arXiv:1410.4447 [hep-ph]].

\bibitem{Adams:2015hiv}
D.~Adams, A.~Arce, L.~Asquith, M.~Backovic, T.~Barillari, P.~Berta, D.~Bertolini, A.~Buckley, J.~Butterworth and R.~C.~Camacho Toro, \textit{et al.}
Eur. Phys. J. C \textbf{75}, no.9, 409 (2015)
[arXiv:1504.00679 [hep-ph]].

\bibitem{ATLAS:2019fgd}
G.~Aad \textit{et al.} [ATLAS],
JHEP \textbf{03}, 145 (2020)
[arXiv:1910.08447 [hep-ex]].

\bibitem{CMS:2020cpy}
A.~M.~Sirunyan \textit{et al.} [CMS],
Eur. Phys. J. C \textbf{80}, no.8, 752 (2020)
[arXiv:2001.10086 [hep-ex]].

\bibitem{CMS:2024nsz}
V.~Chekhovsky \textit{et al.} [CMS],
Rept. Prog. Phys. \textbf{88}, no.6, 067802 (2025)
[arXiv:2412.03747 [hep-ex]].

\bibitem{CMS:2025wfw}
A.~Hayrapetyan \textit{et al.} [CMS],
[arXiv:2503.06726 [hep-ex]].


\bibitem{Mangano:2017plv}
M.~L.~Mangano and B.~Nachman,
Eur. Phys. J. C \textbf{78}, no.4, 343 (2018)
[arXiv:1708.08369 [hep-ph]].

\bibitem{Milhano:2017nzm}
G.~Milhano, U.~A.~Wiedemann and K.~C.~Zapp,
Phys. Lett. B \textbf{779}, 409-413 (2018)
[arXiv:1707.04142 [hep-ph]].

\bibitem{Chang:2017gkt}
N.~B.~Chang, S.~Cao and G.~Y.~Qin,
Phys. Lett. B \textbf{781}, 423-432 (2018)
[arXiv:1707.03767 [hep-ph]].

\bibitem{KunnawalkamElayavalli:2017hxo}
R.~Kunnawalkam Elayavalli and K.~C.~Zapp,
JHEP \textbf{07}, 141 (2017)
[arXiv:1707.01539 [hep-ph]].

\bibitem{Ringer:2019rfk}
F.~Ringer, B.~W.~Xiao and F.~Yuan,
Phys. Lett. B \textbf{808}, 135634 (2020)
[arXiv:1907.12541 [hep-ph]].

\bibitem{Casalderrey-Solana:2019ubu}
J.~Casalderrey-Solana, G.~Milhano, D.~Pablos and K.~Rajagopal,
JHEP \textbf{01}, 044 (2020)
[arXiv:1907.11248 [hep-ph]].

\bibitem{Caucal:2019uvr}
P.~Caucal, E.~Iancu and G.~Soyez,
JHEP \textbf{10}, 273 (2019)
[arXiv:1907.04866 [hep-ph]].

\bibitem{Caucal:2021cfb}
P.~Caucal, A.~Soto-Ontoso and A.~Takacs,
Phys. Rev. D \textbf{105}, no.11, 114046 (2022)
[arXiv:2111.14768 [hep-ph]].

\bibitem{Cunqueiro:2023vxl}
L.~Cunqueiro, D.~Pablos, A.~Soto-Ontoso, M.~Spousta, A.~Takacs and M.~Verweij,
Phys. Rev. D \textbf{110}, no.1, 014015 (2024)
[arXiv:2311.07643 [hep-ph]].

\bibitem{JETSCAPE:2023hqn}
Y.~Tachibana \textit{et al.} [JETSCAPE],
Phys. Rev. C \textbf{110}, no.4, 044907 (2024)
[arXiv:2301.02485 [hep-ph]].

\bibitem{Chien:2024uax}
Y.~T.~Chien, O.~Fedkevych, D.~Reichelt and S.~Schumann,
JHEP \textbf{07}, 230 (2024)
[arXiv:2404.04168 [hep-ph]].

\bibitem{Apolinario:2024equ}
L.~Apolin\'ario, Y.~T.~Chien and L.~Cunqueiro Mendez,
Int. J. Mod. Phys. E \textbf{33}, no.07, 2430003 (2024)

\bibitem{Moraes:2007rq}
A.~Moraes, C.~Buttar and I.~Dawson,
Eur. Phys. J. C \textbf{50}, 435-466 (2007)

\bibitem{Cacciari:2007fd}
M.~Cacciari and G.~P.~Salam,
Phys. Lett. B \textbf{659}, 119-126 (2008)
[arXiv:0707.1378 [hep-ph]].

\bibitem{Larkoski:2017jix}
A.~J.~Larkoski, I.~Moult and B.~Nachman,
Phys. Rept. \textbf{841}, 1-63 (2020)
[arXiv:1709.04464 [hep-ph]].

\bibitem{Kogler:2018hem}
R.~Kogler, B.~Nachman, A.~Schmidt, L.~Asquith, M.~Campanelli, C.~Delitzsch, P.~Harris, A.~Hinzmann, D.~Kar and C.~McLean, \textit{et al.}
Rev. Mod. Phys. \textbf{91}, no.4, 045003 (2019)
[arXiv:1803.06991 [hep-ex]].

\bibitem{Marzani:2019hun}
S.~Marzani, G.~Soyez and M.~Spannowsky,
Lect. Notes Phys. \textbf{958}, pp. (2019)
Springer, 2019,
[arXiv:1901.10342 [hep-ph]].

\bibitem{Chien:2010kc}
Y.~T.~Chien and M.~D.~Schwartz,
JHEP \textbf{08}, 058 (2010)
[arXiv:1005.1644 [hep-ph]].

\bibitem{Chen:2021uws}
Y.~Chen, A.~Badea, A.~Baty, P.~Chang, Y.~T.~Chien, G.~M.~Innocenti, M.~Maggi, C.~McGinn, D.~V.~Perepelitsa and M.~Peters, \textit{et al.}
JHEP \textbf{06}, 008 (2022)
[arXiv:2111.09914 [hep-ex]].

\bibitem{H1:2024pvu}
V.~Andreev \textit{et al.} [H1],
Eur. Phys. J. C \textbf{84}, no.7, 718 (2024)
[arXiv:2403.10134 [hep-ex]].

\bibitem{CDF:2011loy}
T.~Aaltonen \textit{et al.} [CDF],
Phys. Rev. D \textbf{85}, 091101 (2012)
[arXiv:1106.5952 [hep-ex]].

\bibitem{STAR:2021lvw}
M.~Abdallah \textit{et al.} [STAR],
Phys. Rev. D \textbf{104}, no.5, 052007 (2021)
[arXiv:2103.13286 [hep-ex]].

\bibitem{ATLAS:2012am}
G.~Aad \textit{et al.} [ATLAS],
JHEP \textbf{05}, 128 (2012)
[arXiv:1203.4606 [hep-ex]].

\bibitem{CMS:2013kfv}
S.~Chatrchyan \textit{et al.} [CMS],
JHEP \textbf{05}, 090 (2013)
[arXiv:1303.4811 [hep-ex]].

\bibitem{ALICE:2017nij}
S.~Acharya \textit{et al.} [ALICE],
Phys. Lett. B \textbf{776}, 249-264 (2018)
[arXiv:1702.00804 [nucl-ex]].

\bibitem{CMS:2017pcy}
A.~M.~Sirunyan \textit{et al.} [CMS],
Eur. Phys. J. C \textbf{77}, no.7, 467 (2017)
[arXiv:1703.06330 [hep-ex]].

\bibitem{ATLAS:2017zda}
M.~Aaboud \textit{et al.} [ATLAS],
Phys. Rev. Lett. \textbf{121}, no.9, 092001 (2018)
[arXiv:1711.08341 [hep-ex]].

\bibitem{ATLAS:2018jsv}
 [ATLAS],
ATLAS-CONF-2018-014.

\bibitem{CMS:2018fof}
A.~M.~Sirunyan \textit{et al.} [CMS],
JHEP \textbf{10}, 161 (2018)
[arXiv:1805.05145 [hep-ex]].

\bibitem{CMS:2018vzn}
A.~M.~Sirunyan \textit{et al.} [CMS],
JHEP \textbf{11}, 113 (2018)
[arXiv:1807.05974 [hep-ex]].

\bibitem{ALICE:2024jtb}
S.~Acharya \textit{et al.} [ALICE],
Phys. Lett. B \textbf{864}, 139409 (2025)
[arXiv:2411.03106 [nucl-ex]].

\bibitem{LHCb:2025tvf}
R.~Aaij \textit{et al.} [LHCb],
[arXiv:2505.11955 [hep-ex]].

\bibitem{Gehrmann-DeRidder:2007vsv}
A.~Gehrmann-De Ridder, T.~Gehrmann, E.~W.~N.~Glover and G.~Heinrich,
JHEP \textbf{12}, 094 (2007)
[arXiv:0711.4711 [hep-ph]].

\bibitem{Baron:2018nfz}
J.~Baron, S.~Marzani and V.~Theeuwes,
JHEP \textbf{08}, 105 (2018)
[erratum: JHEP \textbf{05}, 056 (2019)]
[arXiv:1803.04719 [hep-ph]].

\bibitem{Kardos:2018kth}
A.~Kardos, G.~Somogyi and Z.~Tr\'ocs\'anyi,
Phys. Lett. B \textbf{786}, 313-318 (2018)
[arXiv:1807.11472 [hep-ph]].

\bibitem{DelDuca:2016ily}
V.~Del Duca, C.~Duhr, A.~Kardos, G.~Somogyi, Z.~Sz\H{o}r, Z.~Tr\'ocs\'anyi and Z.~Tulip\'ant,
Phys. Rev. D \textbf{94}, no.7, 074019 (2016)
[arXiv:1606.03453 [hep-ph]].

\bibitem{Weinzierl:2009ms}
S.~Weinzierl,
JHEP \textbf{06}, 041 (2009)
[arXiv:0904.1077 [hep-ph]].

\bibitem{Dasgupta:2012hg}
M.~Dasgupta, K.~Khelifa-Kerfa, S.~Marzani and M.~Spannowsky,
JHEP \textbf{10}, 126 (2012)
[arXiv:1207.1640 [hep-ph]].

\bibitem{Frye:2016okc}
C.~Frye, A.~J.~Larkoski, M.~D.~Schwartz and K.~Yan,
[arXiv:1603.06375 [hep-ph]].



\bibitem{Idilbi:2016hoa}
A.~Idilbi and C.~Kim,
J. Korean Phys. Soc. \textbf{73}, no.9, 1230-1239 (2018)
[arXiv:1606.05429 [hep-ph]].


\bibitem{Kardos:2020gty}
A.~Kardos, A.~J.~Larkoski and Z.~Tr\'ocs\'anyi,
Phys. Lett. B \textbf{809}, 135704 (2020)
[arXiv:2002.00942 [hep-ph]].

\bibitem{Ziani:2021dxr}
N.~Ziani, K.~Khelifa-Kerfa and Y.~Delenda,
Eur. Phys. J. C \textbf{81}, 570 (2021)
[arXiv:2104.11060 [hep-ph]].



\bibitem{Catani:1990rr}
S.~Catani, B.~R.~Webber and G.~Marchesini,
Nucl. Phys. B \textbf{349}, 635-654 (1991)

\bibitem{Catani:1993hr}
S.~Catani, Y.~L.~Dokshitzer, M.~H.~Seymour and B.~R.~Webber,
Nucl. Phys. B \textbf{406}, 187-224 (1993)

\bibitem{Banfi:2004yd}
A.~Banfi, G.~P.~Salam and G.~Zanderighi,
JHEP \textbf{03}, 073 (2005)
[arXiv:hep-ph/0407286 [hep-ph]].

\bibitem{Li:2012bw}
H.~n.~Li, Z.~Li and C.~P.~Yuan,
Phys. Rev. D \textbf{87}, 074025 (2013)
[arXiv:1206.1344 [hep-ph]].

\bibitem{Chien:2012ur}
Y.~T.~Chien, R.~Kelley, M.~D.~Schwartz and H.~X.~Zhu,
Phys. Rev. D \textbf{87}, no.1, 014010 (2013)
[arXiv:1208.0010 [hep-ph]].

\bibitem{Frye:2016aiz}
C.~Frye, A.~J.~Larkoski, M.~D.~Schwartz and K.~Yan,
JHEP \textbf{07}, 064 (2016)
[arXiv:1603.09338 [hep-ph]].

\bibitem{Kolodrubetz:2016dzb}
D.~W.~Kolodrubetz, P.~Pietrulewicz, I.~W.~Stewart, F.~J.~Tackmann and W.~J.~Waalewijn,
JHEP \textbf{12}, 054 (2016)
[arXiv:1605.08038 [hep-ph]].

\bibitem{Marzani:2017kqd}
S.~Marzani, L.~Schunk and G.~Soyez,
Eur. Phys. J. C \textbf{78}, no.2, 96 (2018)
[arXiv:1712.05105 [hep-ph]].

\bibitem{Kang:2018jwa}
Z.~B.~Kang, K.~Lee, X.~Liu and F.~Ringer,
JHEP \textbf{10}, 137 (2018)
[arXiv:1803.03645 [hep-ph]].

\bibitem{Balsiger:2019tne}
M.~Balsiger, T.~Becher and D.~Y.~Shao,
JHEP \textbf{04}, 020 (2019)
[arXiv:1901.09038 [hep-ph]].

\bibitem{Gaid:2024tie}
S.~Gaid, Y.~Delenda and R.~Soualah,
JHEP \textbf{12}, 179 (2024)
[arXiv:2409.12260 [hep-ph]].

\bibitem{Hoang:2025uaa}
A.~H.~Hoang, V.~Mateu, M.~D.~Schwartz and I.~W.~Stewart,
[arXiv:2506.09130 [hep-ph]].

\bibitem{Dasgupta:2001sh}
M.~Dasgupta and G.~P.~Salam,
Phys. Lett. B \textbf{512}, 323-330 (2001)
[arXiv:hep-ph/0104277 [hep-ph]].

\bibitem{Dasgupta:2002bw}
M.~Dasgupta and G.~P.~Salam,
JHEP \textbf{03}, 017 (2002)
[arXiv:hep-ph/0203009 [hep-ph]].

\bibitem{DuranDelgado:2011tp}
R.~M.~Duran Delgado, J.~R.~Forshaw, S.~Marzani and M.~H.~Seymour,
JHEP \textbf{08}, 157 (2011)
[arXiv:1107.2084 [hep-ph]].

\bibitem{Khelifa-Kerfa:2011quw}
K.~Khelifa-Kerfa,
JHEP \textbf{02}, 072 (2012)
[arXiv:1111.2016 [hep-ph]].

\bibitem{Schwartz:2014wha}
M.~D.~Schwartz and H.~X.~Zhu,
Phys. Rev. D \textbf{90}, no.6, 065004 (2014)
[arXiv:1403.4949 [hep-ph]].

\bibitem{Larkoski:2015zka}
A.~J.~Larkoski, I.~Moult and D.~Neill,
JHEP \textbf{09}, 143 (2015)
[arXiv:1501.04596 [hep-ph]].

\bibitem{Larkoski:2016zzc}
A.~J.~Larkoski, I.~Moult and D.~Neill,
JHEP \textbf{11}, 089 (2016)
[arXiv:1609.04011 [hep-ph]].

\bibitem{Becher:2016mmh}
T.~Becher, M.~Neubert, L.~Rothen and D.~Y.~Shao,
JHEP \textbf{11}, 019 (2016)
[erratum: JHEP \textbf{05}, 154 (2017)]
[arXiv:1605.02737 [hep-ph]].

\bibitem{Becher:2017nof}
T.~Becher, R.~Rahn and D.~Y.~Shao,
JHEP \textbf{10}, 030 (2017)
[arXiv:1708.04516 [hep-ph]].

\bibitem{Caletti:2021oor}
S.~Caletti, O.~Fedkevych, S.~Marzani, D.~Reichelt, S.~Schumann, G.~Soyez and V.~Theeuwes,
JHEP \textbf{07}, 076 (2021)
[arXiv:2104.06920 [hep-ph]].

\bibitem{Becher:2021zkk}
T.~Becher, M.~Neubert and D.~Y.~Shao,
Phys. Rev. Lett. \textbf{127}, no.21, 212002 (2021)
[arXiv:2107.01212 [hep-ph]].

\bibitem{Becher:2023vrh}
T.~Becher, N.~Schalch and X.~Xu,
Phys. Rev. Lett. \textbf{132}, no.8, 8 (2024)
[arXiv:2307.02283 [hep-ph]].

\bibitem{Becher:2023mtx}
T.~Becher, M.~Neubert, D.~Y.~Shao and M.~Stillger,
JHEP \textbf{12}, 116 (2023)
[arXiv:2307.06359 [hep-ph]].

\bibitem{Korchemsky:1999kt}
G.~P.~Korchemsky and G.~F.~Sterman,
Nucl. Phys. B \textbf{555}, 335-351 (1999)
[arXiv:hep-ph/9902341 [hep-ph]].

\bibitem{Dasgupta:2007wa}
M.~Dasgupta, L.~Magnea and G.~P.~Salam,
JHEP \textbf{02}, 055 (2008)
[arXiv:0712.3014 [hep-ph]].

\bibitem{Jouttenus:2013hs}
T.~T.~Jouttenus, I.~W.~Stewart, F.~J.~Tackmann and W.~J.~Waalewijn,
Phys. Rev. D \textbf{88}, no.5, 054031 (2013)
[arXiv:1302.0846 [hep-ph]].

\bibitem{Stewart:2014nna}
I.~W.~Stewart, F.~J.~Tackmann and W.~J.~Waalewijn,
Phys. Rev. Lett. \textbf{114}, no.9, 092001 (2015)
[arXiv:1405.6722 [hep-ph]].

\bibitem{Reichelt:2021svh}
D.~Reichelt, S.~Caletti, O.~Fedkevych, S.~Marzani, S.~Schumann and G.~Soyez,
JHEP \textbf{03}, 131 (2022)
[arXiv:2112.09545 [hep-ph]].

\bibitem{Ellis:2009su}
S.~D.~Ellis, C.~K.~Vermilion and J.~R.~Walsh,
Phys. Rev. D \textbf{80}, 051501 (2009)
[arXiv:0903.5081 [hep-ph]].

\bibitem{Ellis:2009me}
S.~D.~Ellis, C.~K.~Vermilion and J.~R.~Walsh,
Phys. Rev. D \textbf{81}, 094023 (2010)
[arXiv:0912.0033 [hep-ph]].

\bibitem{Krohn:2009th}
D.~Krohn, J.~Thaler and L.~T.~Wang,
JHEP \textbf{02}, 084 (2010)
[arXiv:0912.1342 [hep-ph]].

\bibitem{Dasgupta:2013ihk}
M.~Dasgupta, A.~Fregoso, S.~Marzani and G.~P.~Salam,
JHEP \textbf{09}, 029 (2013)
[arXiv:1307.0007 [hep-ph]].

\bibitem{Larkoski:2014wba}
A.~J.~Larkoski, S.~Marzani, G.~Soyez and J.~Thaler,
JHEP \textbf{05}, 146 (2014)
[arXiv:1402.2657 [hep-ph]].

\bibitem{Dasgupta:2015lxh}
M.~Dasgupta, L.~Schunk and G.~Soyez,
JHEP \textbf{04}, 166 (2016)
[arXiv:1512.00516 [hep-ph]].

\bibitem{Marzani:2017mva}
S.~Marzani, L.~Schunk and G.~Soyez,
JHEP \textbf{07}, 132 (2017)
[arXiv:1704.02210 [hep-ph]].

\bibitem{Hoang:2017kmk}
A.~H.~Hoang, S.~Mantry, A.~Pathak and I.~W.~Stewart,
Phys. Rev. D \textbf{100}, no.7, 074021 (2019)
[arXiv:1708.02586 [hep-ph]].

\bibitem{Kang:2018vgn}
Z.~B.~Kang, K.~Lee, X.~Liu and F.~Ringer,
Phys. Lett. B \textbf{793}, 41-47 (2019)
[arXiv:1811.06983 [hep-ph]].

\bibitem{Hoang:2019ceu}
A.~H.~Hoang, S.~Mantry, A.~Pathak and I.~W.~Stewart,
JHEP \textbf{12}, 002 (2019)
[arXiv:1906.11843 [hep-ph]].

\bibitem{Ferdinand:2023vaf}
A.~Ferdinand, K.~Lee and A.~Pathak,
Phys. Rev. D \textbf{108}, no.11, L111501 (2023)
[arXiv:2301.03605 [hep-ph]].

\bibitem{Pathak:2023sgi}
A.~Pathak,
JHEP \textbf{08}, 054 (2023)
[arXiv:2301.05714 [hep-ph]].

\bibitem{Dhani:2024gtx}
P.~K.~Dhani, O.~Fedkevych, A.~Ghira, S.~Marzani and G.~Soyez,
JHEP \textbf{02}, 046 (2025)
[arXiv:2410.05415 [hep-ph]].

\bibitem{Hou:2019efy}
T.~J.~Hou, J.~Gao, T.~J.~Hobbs, K.~Xie, S.~Dulat, M.~Guzzi, J.~Huston, P.~Nadolsky, J.~Pumplin and C.~Schmidt, \textit{et al.}
Phys. Rev. D \textbf{103}, no.1, 014013 (2021)
[arXiv:1912.10053 [hep-ph]].

\bibitem{GPE1}
Sacks, J., Welch, W., Mitchell, T. \& Wynn, H. Design and Analysis of Computer Experiments. {\em Statistical Science}. \textbf{4}, 409-423 (1989), http://www.jstor.org/stable/2245858

\bibitem{GPE2}
Rasmussen, C. \& Williams, C. Gaussian Processes for Machine Learning. (The MIT Press,2005,11), https://doi.org/10.7551/mitpress/3206.001.0001

\bibitem{Goodman:2010dyf}
J.~Goodman and J.~Weare,
Commun. Appl. Math. Comput. Sc. \textbf{5}, no.1, 65-80 (2010)

\bibitem{Foreman-Mackey:2012any}
D.~Foreman-Mackey, D.~W.~Hogg, D.~Lang and J.~Goodman,
Publ. Astron. Soc. Pac. \textbf{125}, 306-312 (2013)
[arXiv:1202.3665 [astro-ph.IM]].

\end{thebibliography}
\end{document}